\documentclass[10pt,twocolumn,oneside,final]{IEEEtran}

\usepackage{cite}
\usepackage{graphicx}
\usepackage{amsmath}
\usepackage{times}
\usepackage{latexsym}
\usepackage{bm}
\usepackage{amssymb}
\usepackage[center]{caption2}
\usepackage{array}
\usepackage{fancyhdr}
\usepackage{cite,graphicx,amsmath,amssymb}
\usepackage{citesort}
\usepackage{psfrag}
\usepackage{multirow}

\ifCLASSINFOpdf

\else

\fi

\usepackage[dvipsnames,usenames]{color}
\usepackage[usenames,dvipsnames]{color}

\newcommand{\bfbl}[1]{\color{black}#1}

\newcommand{\bl}[1]{\color{black}#1}

\newtheorem{theorem}{Theorem}
\newtheorem{lemma}{Lemma}

\newtheorem{prop}{Proposition}
\newtheorem{alg}{Algorithm}

\newcommand{\tr}{\text{tr}}

\begin{document}
\title{Linear Precoding for the MIMO Multiple Access Channel
with Finite Alphabet Inputs and \\ Statistical CSI}

\author{Yongpeng Wu, \IEEEmembership{Member, IEEE}, Chao-Kai Wen, \IEEEmembership{Member, IEEE},
Chengshan Xiao, \IEEEmembership{Fellow, IEEE}, \\
Xiqi Gao, \IEEEmembership{Senior Member, IEEE}, and Robert Schober, \IEEEmembership{Fellow, IEEE}.

\thanks{This paper was presented in part at IEEE ICC 2014.}

\thanks{The work of Y. Wu and R. Schober was supported
by the Alexander von Humboldt Foundation. The work of Y. Wu and X. Gao was also supported in part by
 National Natural Science Foundation of China under Grants 61320106003 and 61222102,
the China High-Tech 863 Plan under Grant 2012AA01A506, National Science and Technology Major Project of China under Grants 2013ZX03003004
and 2014ZX03003006-003, and the Program for Jiangsu Innovation Team. The work of C.-K. Wen was supported in part by the MOST of Taiwan
under Grant MOST103-2221-E-110-029-MY3.
 The work of C. Xiao was supported in part by National Science
Foundation under Grants CCF-0915846 and ECCS-1231848.
Part of this
work was carried out while C. Xiao was a visiting professor
at Universit\"{a}t Erlangen-N\"{u}rnberg.
}

\thanks{Y. Wu and R. Schober are with Institute for Digital Communications, Universit\"{a}t Erlangen-N\"{u}rnberg,
Cauerstrasse 7, D-91058 Erlangen, Germany (Email: yongpeng.wu@lnt.de; schober@lnt.de). Y. Wu was with
the National Mobile Communications Research Laboratory, Southeast University, Nanjing, 210096, P. R. China (Email: ypwu@seu.edu.cn).}

\thanks{C.-K. Wen is with the Institute of Communications Engineering, National Sun Yat-sen University, Kaohsiung 804,
Taiwan (Email: chaokaiwen@gmail.com).}

\thanks{C. Xiao is with the Department of Electrical and Computer Engineering,
Missouri University of Science and Technology, Rolla, MO 65409, USA (Email: xiaoc@mst.edu). }

\thanks{X. Gao is with the National Mobile Communications Research Laboratory,
Southeast University, Nanjing, 210096, P. R. China (Email: xqgao@seu.edu.cn). }

}

\maketitle

\begin{abstract}
In this paper, we investigate the design of linear precoders for the multiple-input multiple-output (MIMO) multiple access channel (MAC). We assume that statistical channel state information (CSI) is available at the transmitters and consider the problem under the practical finite alphabet input assumption. First, we derive an asymptotic (in the large system limit) expression for the
weighted sum rate (WSR) of the MIMO MAC with finite alphabet inputs and Weichselberger's MIMO channel model.  Subsequently, we obtain the optimal structures of the linear precoders of the users maximizing the asymptotic WSR and an iterative algorithm for determining the precoders. We show that the complexity of the proposed precoder design is significantly
lower than that of MIMO MAC precoders designed for finite alphabet inputs and instantaneous CSI.
Simulation results for finite alphabet signalling  indicate that the proposed precoder achieves significant performance gains over  existing precoder designs.

\end{abstract}

\begin{keywords}
Finite alphabet, linear precoding, MIMO MAC, statistical CSI.
\end{keywords}

%\newpage

\section{Introduction}
In recent years, the channel capacity and the design of optimum
transmission strategies for the multiple-input multiple-output (MIMO) multiple access channel (MAC)
have been widely studied \cite{goldsmith2003capacity,Yu2002,Yu2004TIT}.  For instance, it was proved in \cite{goldsmith2003capacity} that the boundary of the MIMO MAC capacity region
is achieved by Gaussian input signals. It was further demonstrated in \cite{Yu2002} that, for the MIMO MAC, the optimization of the transmit signal covariance matrices for  weighted sum rate (WSR)
maximization leads to a convex optimization problem. For sum rate maximization, the authors of \cite{Yu2004TIT} developed an efficient
iterative water-filling algorithm for finding the optimal input signal covariance matrices for all users.

However, the results in \cite{goldsmith2003capacity,Yu2002,Yu2004TIT} rely on the critical assumption
of Gaussian input signals.  Although Gaussian inputs are optimal in theory, they are rarely used in practice. Rather, it is well-known that practical communication signals usually are drawn from finite constellation sets,
such as pulse amplitude modulation (PAM), phase shift keying (PSK) modulation, and quadrature amplitude modulation (QAM). These finite constellation sets
differ significantly from the Gaussian idealization \cite{Lozano2006TIT,Xiao2011TSP,Wang2011,Wu2012TVT,Wu2012TWC,Wu2013TCOM}.  Accordingly,  transmission schemes
designed based on the Gaussian input assumption may result in substantial performance losses when finite alphabet inputs are used for transmission.
In \cite{Xiao2011TSP}, the globally optimal linear precoder
design for point-to-point communication systems with finite alphabet inputs was obtained, building upon
earlier works \cite{Xiao2008GlobalCom,Xiao2009ICC,Payaro2009,Lamarca2009}.
For the case of the two-user single-input single-output MAC with finite alphabet inputs, the optimal angle of rotation and the optimal power division
between the transmit signals were found in \cite{Harshan2010TIT} and \cite{harshan2013novel}, respectively.
For the MIMO MAC with an arbitrary number
of users and generic antenna configurations,
 an iterative algorithm for searching for the optimal precoding matrices of all users was proposed in \cite{Wang2011}.

The transmission schemes in \cite{Xiao2011TSP,Wang2011,Wu2012TVT,Wu2012TWC,Wu2013TCOM,Xiao2008GlobalCom,Xiao2009ICC,Payaro2009,Lamarca2009,Harshan2010TIT,harshan2013novel}
require accurate \textit{instantaneous} channel state information (CSI)
at the transmitters for precoder design.
If the channels vary relatively slowly, in frequency division duplex
systems,
the instantaneous CSI can be estimated accurately at
the receiver via uplink training
and then sent to the transmitters through
dedicated feedback links, and in time division duplex systems, the instantaneous CSI can
be obtained by exploiting the reciprocity of uplink and downlink.
Nevertheless, when the mobility of the users increases
and channel fluctuations vary more rapidly, the round-trip delays of the CSI
become non-negligible with respect to
the coherence time of the channels.  In this case,
the obtained instantaneous CSI at the transmitters might be outdated. Therefore, for these scenarios, it is more
reasonable to exploit the channel \textit{statistics} at the transmitter for precoder design,
as the statistics change much more slowly
than the instantaneous channel parameters.

Transmitter design for statistical CSI has received
much attention for the case of Gaussian input signals \cite{tulino2006capacity,Gao,bjornson2010cooperative,wen2011sum,Romain2011TIT,JWang2012TSP,Wu2014TSP}.
For finite alphabet inputs, point-to-point systems,
an efficient precoding algorithm for maximization of the ergodic capacity over Kronecker fading channels was developed in \cite{zeng2012linear}.
Also, in \cite{wen2007asymptotic}, asymptotic (in the large system limit) expressions for the mutual information
of the MIMO MAC with Kronecker fading were derived.  Recently, an iterative algorithm for precoder optimization
for sum rate maximization of the MIMO MAC with Kronecker fading was proposed in \cite{Girnyk2014TWC}.
Despite these previous works, the study of the MIMO MAC
with statistical CSI at the transmitter and finite alphabet inputs remains incomplete, for three reasons: First, the
Kronecker fading model characterizes the correlations of the transmit and the receive antennas separately, which is often not in agreement with
measurements \cite{ozcelik2003deficiencies,weichselberger2006stochastic}.  In contrast,  jointly-correlated fading models, such as
 Weichselberger's model \cite{weichselberger2006stochastic}, do not only account for the correlations at both ends of the link, but
also characterize their mutual dependence. As a consequence,  Weichselberger's model
provides a more general representation of MIMO channels. Second, explicit structures of
the optimal precoders for the MIMO MAC with statistical CSI and finite alphabet inputs have not been reported yet. Third,
in contrast to the sum rate,
the weighted sum rate (WSR) enables service differentiation
in practical communication systems \cite{Christensen2008TWC}.
Thus, it is of interest to study the WSR optimization problem.

In this paper, we investigate the linear precoder design for the $K$-user MIMO MAC assuming Weichselberger's fading model, finite alphabet inputs, and
availability of statistical CSI at the transmitter. By exploiting a random matrix theory tool from statistical physics, referred to as the replica method\footnote{We note
that the replica method has been applied to communications problems before \cite{wen2007asymptotic,Tanaka2002TIT,RMuller2008JSAC,Guo2005TIT}.}, we first derive an
asymptotic expression for the WSR of the MIMO MAC for Weichselberger's fading model in the large system regime where  the numbers of transmit and receive antennas
are both large. The derived expression indicates that the WSR can be obtained asymptotically by calculating the mutual information of each user separately over
equivalent deterministic channels. This property significantly reduces the computational effort for calculation of the WSR.
Furthermore, we prove that the optimal left singular matrix of each user's optimal precoder
corresponds to the eigenmatrix of the transmit correlation
matrix of the user. This result facilitates the derivation of an iterative
algorithm\footnote{It is noted that although we derive the asymptotic WSR in the large system regime,
the proposed algorithm can also be applied for systems with a finite number of antennas.} for computing the optimal precoder for each user.
The proposed algorithm updates the power allocation matrix
 and the right singular matrix of each user in an alternating manner along the gradient decent direction.
We show that the proposed algorithm does not only provide a systematic precoder design method for the MIMO MAC
with  statistical CSI at the transmitter, but also reduces the implementation complexity
by \emph{several orders of magnitude} compared to the precoder design for instantaneous CSI.
Moreover, precoders designed for statistical CSI
can be updated much less frequently than precoders designed for instantaneous CSI as the channel statistics change very slowly compared
to the instantaneous CSI.
Numerical results demonstrate that the proposed design  provides substantial performance gains over
systems without precoding and systems employing precoders designed under the Gaussian input assumption for both  systems with
a moderate number of antennas and  massive MIMO systems \cite{Marzetta2010TWC}.

The remainder of this paper is organized as follows. Section II describes the  considered MIMO MAC model.
In Section III, we derive the asymptotic mutual information expression for the MIMO MAC with statistical CSI at the transmitter and finite alphabet inputs.
 In Section IV, we obtain a closed-form expression for the left singular matrix of
each user's precoder maximizing the asymptotic WSR and propose
an iterative algorithm for determining the precoders of all users.
Numerical results are provided in Section V and our main results are summarized in Section VI.

The following notations are adopted throughout the paper:  Column vectors are represented by lower-case bold-face letters,
and matrices are represented by upper-case bold-face letters. Superscripts $(\cdot)^{T}$, $(\cdot)^{*}$, and $(\cdot)^{H}$
stand for the matrix/vector transpose, conjugate, and conjugate-transpose operations, respectively.
$\rm{det}(\cdot)$ and $\rm{tr}(\cdot)$  denote the matrix determinant and trace operations, respectively. ${\rm{diag}}\left\{\bf{b}\right\}$ and ${\rm{blockdiag}}\left\{{\bf{A}}_k\right\}_{k=1}^{K}$   denote
a diagonal matrix and a block diagonal matrix containing
in the main diagonal and in the block diagonal the elements of vector $\bf{b}$ and matrices ${\bf{A}}_k, k=1,2,\cdots,K$, respectively.
 ${\rm{diag}}\left\{\bf{B}\right\}$  denotes a diagonal matrix containing in the main diagonal
the diagonal elements of matrix $\mathbf{B}$.  $\odot$ and $\bigotimes$ denote the element-wise product
and the Kronecker product of two matrices, respectively. ${\rm vec} \left(\mathbf{A}\right) $ is a column vector which
contains the stacked columns of matrix $\mathbf{A}$.
$[\mathbf{A}]_{mn}$ denotes the element in the
$m$th row and $n$th column of matrix $\mathbf{A}$.
$\left\| {\mathbf{X}} \right\|_F$ denotes the Frobenius norm of matrix $\mathbf{X}$.
${\mathbf{I}}_M$ denotes an $M \times M$ identity matrix,
and $E_V\left[\cdot \right]$ represents the expectation with respect to random variable $V$, which can be a scalar, vector, or matrix.
Finally, $D \mathbf{A}$ denotes the integral measure for the real and imaginary parts of the elements of $\mathbf{A}$. That is, for an $n \times m$ matrix $\mathbf{A}$, we have $D \mathbf{A} = \prod_{i=1}^n \prod_{j=1}^m
\frac{d {\rm{Re}}{[\mathbf{A}]_{ij}} d {\rm{Im}}{[\mathbf{A}]_{ij}} }{\pi}$, where $\rm{Re}$ and $\rm{Im}$ extract the real and imaginary parts, respectively.

\section{System Model}\label{sec:model}
Consider  a single-cell MIMO MAC system with $K$ independent users. We suppose each of the $K$
users has $N_t$ transmit antennas\footnote{For ease of notation, we only consider the case where all users have
the same number of transmit antennas. Note that all results in this paper can be
easily extended to the case when this restriction does not hold.} and the receiver has $N_r$ antennas.  Then, the received signal
$\mathbf{y} \in \mathbb{C}^{N_r \times 1}$ is given by
\begin{equation}\label{model}
{\bf{y}} = \sum\limits_{k = 1}^K {{\bf{H}}_k {\bf{x}}_k }  + {\bf{v}}
\end{equation}%
where ${\mathbf{x}_k} \in \mathbb{C}^{N_t \times 1}$ and ${\mathbf{H}}_k \in \mathbb{C}^{N_r \times N_t}$ denote the transmitted signal
and the channel  matrix of user $k$, respectively.
${\mathbf{v}} \in \mathbb{C}^{N_r \times 1}$ is a zero-mean complex Gaussian noise vector
with covariance matrix\footnote{To simplify our notation, in this paper, without loss of generality, we
normalize the power of the noise to unity.} $\mathbf{I}_{N_r}$.
Furthermore, we make the common assumption (as, e.g., \cite{Soysal2007JSAC,wen2011sum})
that the receiver has the instantaneous CSI of all users, and each transmitter has the statistical CSI of all users.

The transmitted signal vector ${\mathbf{x}_k}$ can be expressed as
\begin{equation}\label{x_signal}
{\mathbf{x}_k} = {{\mathbf{B}}_k {\mathbf{d}}_k }
\end{equation}%
where ${\mathbf{B}}_k \in \mathbb{C}^{N_t \times N_t}$ and ${\mathbf{d}}_k \in \mathbb{C}^{N_t \times 1}$ denote the linear precoding matrix and the
input data vector of user $k$, respectively. Furthermore, we assume ${\mathbf{d}}_k$ is a zero-mean
vector with covariance matrix $\mathbf{I}_{N_t}$. Instead of employing the traditional assumption
of a Gaussian transmit signal, here we assume ${\mathbf{d}}_k$ is taken from
a discrete constellation, where all elements of the constellation are equally likely.
In addition, the transmit signal ${\mathbf{x}_k}$
conforms to the power constraint
\begin{equation}\label{x_constraint_2}
 E_{\mathbf{x}_k}\left[ {\mathbf{x}_k^H {\mathbf{x}_k}} \right] = {{\rm{tr}}\left( {{\mathbf{B}}_k {\mathbf{B}}_k^H} \right)} \leq P_k, \ k = 1,2,\cdots,K.
\end{equation}%

For the jointly-correlated fading MIMO channel, we adopt Weichselberger's model \cite{weichselberger2006stochastic}
throughout this paper, which is also referred to as the unitary--independent--unitary model \cite{Tulino2005TIT}. \
This model jointly characterizes the correlation
at the transmitter and receiver side.  In particular, for user $k$,
$\mathbf{H}_k$ is modeled as
\begin{equation}\label{H_channel}
{\bf{H}}_k  = {\bf{U}}_{{\rm{R}}_k } \left( {{\bf{ \widetilde{G}}}_k  \odot {\bf{W}}_k } \right){\bf{U}}_{{\rm{T}}_k }^H
\end{equation}%
where ${\bf{U}}_{{\rm{R}}_k }  = \left[ {{\bf{u}}_{{\rm{R}}_k ,1} ,{\bf{u}}_{{\rm{R}}_k ,2} , \cdots ,{\bf{u}}_{{\rm{R}}_k ,N_r} } \right] \in \mathbb{C}^{N_r \times N_r}  $ and ${\bf{U}}_{{\rm{T}}_k }  = \left[ {{\bf{u}}_{{\rm{T}}_k ,1} ,{\bf{u}}_{{\rm{T}}_k ,2} , \cdots ,{\bf{u}}_{{\rm{T}}_k ,N_t} } \right] \in \mathbb{C}^{N_t \times N_t}$ represent
deterministic unitary matrices, respectively.  ${\bf{ \widetilde{G}}}_k \in \mathbb{C}^{N_r \times N_t} $ is a  deterministic matrix with real-valued nonnegative elements,
and ${\bf{W}}_k \in \mathbb{C}^{N_r \times N_t}$ is a random matrix with independent identically distributed (i.i.d.) Gaussian elements with zero-mean and unit variance.
We define  ${\bf{G}}_k  = {\bf{\widetilde{{G}}}}_k  \odot {\bf{ \widetilde{{G}}}}_k$ and let $g_{k,n,m}$ denote the $(n,m)$th element of matrix
${\bf{G}}_k$. Here, $\mathbf{G}_k$ is referred to as the ``coupling matrix"  as $g_{k,n,m}$ corresponds to the average coupling energy
between ${\bf{u}}_{{\rm{R}}_k ,n}$ and ${\bf{u}}_{{\rm{T}}_k ,m}$\cite{weichselberger2006stochastic}.  The transmit and
receive correlation matrices of user $k$ can be written as
\begin{equation}\label{correlation}
\begin{array}{l}
 {\bf{R}}_{t,k}  = E_{{\bf{H}}_k}\left[ {{\bf{H}}_k^H {\bf{H}}_k } \right] = {\bf{U}}_{{\rm{T}}_k } \boldsymbol{\Gamma} _{{\rm{T}}_k } {\bf{U}}_{{\rm{T}}_k }^H  \\
 {\bf{R}}_{r,k}  = E_{{\bf{H}}_k}\left[ {{\bf{H}}_k {\bf{H}}_k^H } \right] = {\bf{U}}_{{\rm{R}}_k } \boldsymbol{\Gamma} _{{\rm{R}}_k } {\bf{U}}_{{\rm{R}}_k }^H  \\
 \end{array}
\end{equation}
where  $\boldsymbol{\Gamma}_{{\rm{T}}_k } $ and  $\boldsymbol{\Gamma}_{{\rm{R}}_k }$ are diagonal matrices with main diagonal elements
$\left[ {{\bf{\Gamma }}_{{\rm{T}}_k } } \right]_{mm}  = \sum\nolimits_{n = 1}^{N_r } {g_{k,n,m} }$, $m = 1,2,\cdots,N_t$, and $
\left[ {{\bf{\Gamma }}_{{\rm{R}}_k } } \right]_{nn}  = \sum\nolimits_{m = 1}^{N_t } {g_{k,n,m} }$, $n = 1,2,\cdots,N_r$, respectively.

We note that (\ref{H_channel}) is a general model which includes many popular statistical fading  models as special
cases. For example, if $\mathbf{G}_k$ is a rank-one matrix, then (\ref{H_channel}) reduces to the separately-correlated
Kronecker model \cite{shiu2000fading,xiao2004TWC}. On the other hand, if  ${\bf{U}}_{{\rm{R}}_k }$ and ${\bf{U}}_{{\rm{T}}_k }$  are discrete Fourier transform
matrices, (\ref{H_channel}) corresponds to the virtual channel
representation for uniform linear arrays\cite{sayeed2002deconstructing}.

 We emphasize that  Weichselberger's model avoids the separability assumption of the Kronecker model and can account for arbitrary coupling between the
transmitter and receiver ends. Therefore,  Weichselberger's model improves the capability to correctly model actual MIMO channels.
For example, \cite[Fig. 6]{weichselberger2006stochastic} shows that
Weichselberger's model can provide significantly more accurate estimates of the
mutual information for actual MIMO channels than the Kronecker model.
This was the  motivation for using  Weichselberger's model in previous work, e.g. \cite{Gao,wen2011sum},
and is also the main reason for using it in this paper. Hence, with this flexible model, we can obtain a more accurate theoretical
analysis and more realistic performance results for {practical} communication systems compared to the simple Kronecker model.

\section{Asymptotic WSR of the MIMO MAC with Finite Alphabet Inputs}
We divide all users into two groups, denoted as set $\cal A$ and its
complement set ${\cal{A}}^c$: ${\cal {A}} =\{i_1, i_2, \cdots, i_{K_1} \}\subseteq \{1,2, \cdots, K\}$ and ${\cal {A}}^c = \{j_1, j_2, \cdots, j_{K_2}\}$,
$K_1+K_2= K$. Also, we define ${\bf{H}}_{\cal {A}} = \left[ {{\bf{H}}_{i_1} \ {\bf{H}}_{i_2}  \cdots \mathbf{H}_{i_{K_1}} } \right]$,
$\mathbf{d}_{\cal {A}} = \left[ \mathbf{d}_{i_1}^T \ \mathbf{d}_{i_2}^T  \cdots \mathbf{d}_{i_{K_1}}^T \right]^T$,
$ \mathbf{d}_{{\cal {A}}^c} = \left[ \mathbf{d}_{j_1}^T \ \mathbf{d}_{j_2}^T  \cdots \mathbf{d}_{j_{K_2}}^T \right]^T$,
$\mathbf{B}_{\cal {A}} = {\rm{blockdiag}} \left\{ \mathbf{B}_{i_1},\mathbf{B}_{i_2}, \cdots, \mathbf{B}_{i_{K_1}} \right\}$,
and ${\bf{y}}_{\cal {A}} = {\bf{H}}_{\cal {A}} \mathbf{B}_{\cal {A}} \mathbf{d}_{\cal {A}} + {\bf{v}}$. Then, the achievable rate region  $(R_1, R_2, \cdots, R_K)$ of the $K$-user MIMO MAC satisfies the following
conditions\cite{Cover}:
\begin{equation}\label{capacity_region}
    \sum_{i\in {\cal A}} R_i \leq I
    \left( \mathbf{d}_{\cal A}; \mathbf{y} | \mathbf{d}_{{\cal {A}}^c} \right), \quad \forall {\cal {A}} \subseteq \{ 1, 2, \cdots, K \}
\end{equation}%
where
\begin{equation}\label{mutual_info_1}
I\left( {{\bf{d}}_{\cal {A}} ;{\bf{y}}\left| {{\bf{d}}_{{\cal A}^c } } \right.} \right) \!\! = \!\! E_{{\bf{H}}_{\cal {A}} }\! \left[ \! {E_{{\bf{d}}_{\cal {A}} ,{\bf{y}}_{\cal {A}} } \! \left[ \!{\log _2 \frac{{p\left( {{\bf{y}}_{\cal {A}} \left| {{\bf{d}}_{\cal {A}} {\bf{,H}}_{\cal {A}} } \right.} \right)}}{{p\left( {{\bf{y}}_{\cal {A}} \left| {{\bf{H}}_{\cal {A}} } \right.} \right)}}\left| {\bf{H}} \right._{\cal {A}} } \! \right] }  \right].
\end{equation}%
In (\ref{mutual_info_1}), ${p} ({\bf{y}}_{\cal A} |{\bf{H}}_{\cal A})$ denotes the marginal probability density function (p.d.f.) of $p(\mathbf{d}_{\cal A},\mathbf{y}_{\cal A}|\mathbf{H}_{\cal A})$. As a result, we have
\begin{multline}\label{eq:Finite_Mutual}
I (\mathbf{d}_{\cal A};\mathbf{y} \left| {{\bf{d}}_{{{\cal A}^c} } } \right. )   =  \\
 - E_{\mathbf{H}_{\cal A}}\left[ E_{\mathbf{y}_{\cal A}} \left[\left.\log_2 E_{\mathbf{d}_{\cal A}}\left[e^{-\left\|\mathbf{y}_{\cal A}-\mathbf{H}_{\cal A}  \mathbf{B}_{\cal A} \mathbf{d}_{\cal A}\right\|^2}\right]\right|\mathbf{H}_{\cal A} \right]  \right] \\
- N_r \log_2 e.
\end{multline}%
The expectation in (\ref{eq:Finite_Mutual}) can be evaluated numerically by Monte-Carlo simulation. However,
for a large number of antennas, the associated computational complexity could be enormous. Therefore,
by employing the replica method, a classical technique from statistical physics, we obtain an asymptotic expression
for (\ref{eq:Finite_Mutual}) as detailed in the following.

\subsection{Some Useful Definitions}
We first introduce some useful definitions. Consider a virtual\footnote{The virtual channel model does not relate to a physical channel but it plays an important role in the
derivation of our final asymptotic expression (\ref{eq:GAUMutuall_2}) in Proposition \ref{prop:ach_rate_mac}.} MIMO channel defined by
\begin{equation}\label{eq:EqScalGAUEach}
\mathbf{z}_{\cal A}= \sqrt{\mathbf{T}_{\cal A}} \mathbf{B}_{\cal A} \mathbf{d}_{\cal A} + \check{\bf v}_{\cal A}.
\end{equation}
Hence, $\mathbf{T}_{\cal A} \in \mathbb{C}^{K_1 N_t \times K_1 N_t} $ is
given by $\mathbf{T}_{\cal A} =  {\rm{blockdiag}}$ $\left\{ \mathbf{T}_{i_1}, \mathbf{T}_{i_2}, \cdots, \mathbf{T}_{i_{K_1}} \right\}
\in\mathbb{C}^{K_1 N_t \times K_1 N_t}$, where $\mathbf{T}_{i_k}\in\mathbb{C}^{N_t \times N_t}$ is a
deterministic matrix, $k = 1,2,\cdots,K_1$. $\check{\bf v}_{\cal A} \in \mathbb{C}^{K_1 N_r \times 1} $  is a standard complex Gaussian random vector with i.i.d. elements.
The minimum mean square error (MMSE) estimate of  signal vector $\mathbf{d}_{\cal A}$ given (\ref{eq:EqScalGAUEach}) can be expressed
as
\begin{equation}\label{eq:hatx_k}
 \hat{\mathbf{d}}_{\cal A} = E_{\mathbf{d}_{\cal A}} \left[ \mathbf{d}_{\cal A} \Big|\mathbf{z}_{\cal A},\sqrt{\mathbf{T}_{\cal A}},\mathbf{B}_{\cal A}  \right].
 \end{equation}
 Define the following mean square error (MSE) matrix
  \begin{equation}\label{eq:mse}
    \boldsymbol{\Omega}_{\cal A} = \mathbf{B}_{\cal A} \mathbf{E}_{\cal A} \mathbf{B}_{\cal A}^H
\end{equation}
where
\begin{equation}\label{eq:mse_1}
\mathbf{E}_{\cal A} =  E_{\mathbf{z}_{\cal A}} \left[ E_{\mathbf{d}_{\cal A}}\left[( \mathbf{d}_{\cal A} - \hat{\mathbf{d}}_{\cal A}) (\mathbf{d}_{\cal A} - \hat{\mathbf{d}}_{\cal A})^H \right] \right].
\end{equation}
Define the matrices of the $i_k$th ($i_1 \leq i_k \leq i_{K_1}$) user $\boldsymbol{\Omega}_{i_k}$
and $\boldsymbol{E}_{i_k}$ as  submatrices obtained by extracting the $\left( (k - 1) N_t +
1 \right)$th to the $(k N_t)$th row and column elements of matrices $\boldsymbol{\Omega}_{\cal A}$ and $\mathbf{E}_{\cal A}$,
respectively.

{\bl The component matrices $\mathbf{T}_{i_k}$ in $\mathbf{T}_{\cal A}$ are functions of auxiliary variables $\{ \mathbf{R}_{i_k}, \boldsymbol{\gamma}_{i_k}, \boldsymbol{\psi}_{i_k} \}$, which are the solutions of the following set of coupled equations:
\begin{equation} \label{eq:eqChMatrixTR} \left\{\begin{aligned}
\mathbf{T}_{i_k} & =\mathbf{U}_{{\rm T}_{i_k}}{\rm diag}\left( \mathbf{G}_{i_k}^T \boldsymbol{\gamma}_{i_k} \right)\mathbf{U}_{{\rm T}_{i_k}}^{H}\in\mathbb{C}^{N_t \times N_t}\\
\mathbf{R}_{i_k} &= \mathbf{U}_{{\rm R}_{i_k}}{\rm diag}\left( \mathbf{G}_{i_k} \boldsymbol{\psi}_{i_k}\right)\mathbf{U}_{{\rm R}_{i_k}}^{H}\in\mathbb{C}^{N_r \times N_r}
\end{aligned}\right.
\end{equation}% where we have used $\boldsymbol{\gamma}_{i_k} = [\gamma_{i_k,1}, \gamma_{i_k,2} ,\cdots, \gamma_{i_k,N_r}]^T$, $\boldsymbol{\psi}_{i_k} = [\psi_{i_k,1}, \psi_{i_k,2} ,\cdots, \psi_{i_k,N_t}]^T$
where $\boldsymbol{\gamma}_{i_k} = [\gamma_{i_k,1}, \gamma_{i_k,2} ,\cdots, \gamma_{i_k,N_r}]^T$ and $\boldsymbol{\psi}_{i_k} = [\psi_{i_k,1}, \psi_{i_k,2} ,\cdots, \psi_{i_k,N_t}]^T$ with
\begin{equation}\label{eq:Varsigma_k-MSE}
\left\{\begin{aligned}
\gamma_{i_k,n} &=  \mathbf{u}_{{\rm R}_{i_k},n}^{H} \left(\mathbf{I}_{N_r}+\mathbf{R}_{\cal A}\right)^{-1} \mathbf{u}_{{\rm R}_{i_k},n}, \ n = 1,2,\cdots,N_r \\
\psi_{i_k,m}   &= \mathbf{u}_{{\rm T}_{i_k},m}^{H} \mathbf{\Omega}_{i_k}  \mathbf{u}_{{\rm T}_{i_k},m}, \  m = 1,2,\cdots,N_t
\end{aligned}
\right.
\end{equation}%
and $\mathbf{R}_{\cal A} = \sum_{k=1}^{K_1} \mathbf{R}_{i_k}$. Computing $\mathbf{T}_{i_k}$ requires
finding $\{ \mathbf{R}_{i_k}, \boldsymbol{\gamma}_{i_k}, \boldsymbol{\psi}_{i_k} \}$ through  fixed point
equations (\ref{eq:eqChMatrixTR}) and (\ref{eq:Varsigma_k-MSE}).
 We will show later that in the
asymptotic regime the mutual information in (\ref{eq:Finite_Mutual}) can be evaluated based on the
virtual MIMO channel\footnote{\bl The virtual MIMO channel model in (\ref{eq:EqScalGAUEach}) is only used to evaluate the asymptotic {mutual information} of the actual channel model in (\ref{model}) in the large system regime. Therefore, the dimensionality of the virtual MIMO channel model does not need to be
identical to that of the channel model in (\ref{model}). The dimensionality of the virtual MIMO channel model is detailed in Appendix \ref{sec:proof_ach_rate_mac}. } in (\ref{eq:EqScalGAUEach}). However, in contrast to channel matrix $\mathbf{H}_{\cal A}$ in (\ref{eq:Finite_Mutual}),
for the virtual MIMO channel, channel matrix
$\sqrt{\mathbf{T}_{\cal A}} $ is deterministic.

\subsection{Asymptotic Mutual Information}
Suppose the transmit signal ${\mathbf{d}}_k$ is taken from a discrete constellation with cardinality
$Q_k$. Define $M_k = {Q_k^{N_t}}$, let ${\cal{S}}_k$ denote the constellation set for user $k$, and let
 ${\mathbf{a}}_{k,j}$ denote the $j$th element of ${\cal{S}}_k$, $k =1,2,\cdots ,K$,
 $j = 1,2, \cdots,M_k$. {\bfbl We define the large system limit as the scenario where $N_r$ and $N_t$  are large but the ratio
$\beta = N_t/ N_r$ is fixed. }Now, we are ready to provide a simplified asymptotic expression for (\ref{eq:Finite_Mutual}).

\begin{prop}\label{prop:ach_rate_mac}
For the MIMO MAC model (\ref{model}), in the large system limit the mutual information in (\ref{eq:Finite_Mutual}) can be asymptotically approximated by
\begin{multline}\label{eq:GAUMutuall_2}
I\left( {{\bf{d}}_{\cal A} ;{\bf{y}}\left| {{\bf{d}}_{{{\cal A}^c} } } \right.} \right) \simeq
\sum\limits_{i \in \cal A} {I\left( {{\bf{d}}_{i_k } ;{\bf{z}}_{i_k } \left| {\sqrt {{\bf{T}}_{i_k } } {\bf{B}}_{i_k } } \right.} \right)} \\
 + \log_2 \det\left(\mathbf{I}_{N_r}+\mathbf{R}_{\cal A}\right) - \log_2 e \sum_{k=1}^{K_1} \boldsymbol{\gamma}_{i_k}^T \mathbf{G}_{i_k} \boldsymbol{\psi}_{i_k}
\end{multline}
where
\begin{multline}\label{eq:mutual_info}
I\left( {{\bf{d}}_{i_k } ;{\bf{z}}_{i_k } \left| {\sqrt {{\bf{T}}_{i_k } } {\bf{B}}_{i_k } } \right.} \right) = \log _2 M_{i_k }  - \frac{1}{{M_{i_k } }}  \\
\times \sum\limits_{m = 1}^{M_{i_k } } {E_{{\bf{v}} } \! \left\{ \!  {\log _2 \sum\limits_{p = 1}^{M_{i_k } } {e^{ - \left( {\left\| {\sqrt {{\bf{T}}_{i_k } } {\bf{B}}_{i_k } \left( {{\bf{a}}_{k,p} \!  - \! {\bf{a}}_{k,m} } \right) + {\bf{v}} } \right\|^2  - \left\| {{\bf{v}} } \right\|^2 } \right)} } } \! \right\}}.
\end{multline}

\begin{proof}
See Appendix \ref{sec:proof_ach_rate_mac}.
\end{proof}
\end{prop}

{\bl \textit{Remark 1:}  The  asymptotic expressions provided in Proposition \ref{prop:ach_rate_mac} constitute approximations for matrices of finite dimension.} {\bl In addition, because the derivation
of the asymptotic expression is based on the replica method, wherein some steps lack a rigorous proof, we state the result in a proposition rather than a theorem. }

{\bl \textit{Remark 2:} As mentioned above,  Weichselberger's model is a general channel model. Therefore, the unified expression
in Proposition \ref{prop:ach_rate_mac} is applicable to many special cases.
{\bfbl For example, if $\mathbf{G}_k$ is a rank-one matrix, (\ref{eq:GAUMutuall_2}) reduces to
\cite[Eq. (28)]{Girnyk2014TWC}, which was derived for the Kronecker model.}

{\bl \textit{Remark 3:}
 $\boldsymbol{\gamma}_{i_k}$ and $\boldsymbol{\psi}_{i_k}$ in Proposition \ref{prop:ach_rate_mac} can be obtained
 through the fixed-point equations in (\ref{eq:Varsigma_k-MSE}).  From statistical physics, it is known that
 there are multiple solutions for $\gamma_{i_k,n}$ and $\psi_{i_k,m}$ that satisfy\footnote{This effect is called the phenomenon of phase coexistence.
 For details on this phenomenon, please refer to \cite{Tanaka2002TIT}.} (\ref{eq:Varsigma_k-MSE}).   For the problem at hand, the solution {minimizing}
 $I\left( {{\bf{d}}_{\cal A} ;{\bf{y}}\left| {{\bf{d}}_{{{\cal A}^c} } } \right.} \right)$ in (\ref{eq:GAUMutuall_2}) yields the mutual information.}

{\bfbl Before proceeding, let us recall some notations used in this paper. For the virtual channel model (\ref{eq:EqScalGAUEach}),
the virtual channel matrix and the corresponding asymptotic parameters obtained
for different sets ${\cal A}$ based on the fixed point equations
(\ref{eq:eqChMatrixTR}) and (\ref{eq:Varsigma_k-MSE}) are different due to the equality {$\mathbf{R}_{\cal A} = \sum_{i_k \in {\cal A} }
\mathbf{R}_{i_k}$.} Therefore, we define the set ${\cal A}_k = \left\{1,2,\cdots,k\right\}$. Then, we denote the virtual channel matrix and the corresponding
asymptotic parameters obtained from the fixed point equations (\ref{eq:eqChMatrixTR}) and
(\ref{eq:Varsigma_k-MSE}), (\ref{eq:EqScalGAUEach}), and (\ref{eq:mse}) for ${\cal{A}} = {\cal A}_k$
as $\mathbf{T}^{(k)}_{t}$, $\mathbf{R}^{(k)}_{t}$, ${\boldsymbol{\gamma}}^{(k)}_{t}$,
${\boldsymbol{\psi}}^{(k)}_{t}$, ${\bf{z}}_t^{\left( k \right)}$, ${\boldsymbol{E }}^{(k)}_{t}$, and ${\boldsymbol{\Omega }}^{(k)}_{t}$, $t = 1,2,\cdots,k$.
Here, the indices $[i_1,i_2,\cdots,i_{K_1}]$ in (\ref{eq:eqChMatrixTR}) and (\ref{eq:Varsigma_k-MSE}) are $[1,2,\cdots,k]$, and $t$ refers to the individual
users in the set $[1,2,\cdots,k]$.
}

\section{Linear Precoding Design for the MIMO MAC}\label{sec:Precoding}
{\bfbl In this section, we first formulate the WSR optimization problem  for linear precoder design for the MIMO MAC.
Then, we establish the structure of the asymptotically optimal precoders maximizing the WSR in the large system limit.
Finally, we propose an iterative algorithm for finding the optimal precoders.

\subsection{Weighted Asymptotic Sum Rate}
It is well known that the achievable rate region of the MIMO MAC $(R_1, R_2, \cdots, R_K)$ can be obtained by solving the WSR optimization problem \cite{goldsmith2003capacity}.
Without loss of generality, assume  weights $\mu _1 \geq \mu _2 \geq \cdots \geq  \mu _K \geq \mu _{K+1} = 0$, \emph{i.e.},
 the users are decoded in the order $K, K -1, \cdots, 1$ \cite{Wang2011}.  Then, the WSR problem can be expressed as {
\begin{equation} \label{eq:obj_wsr_2}
\begin{aligned}
    { {\rm WSR}} =
    \mathop {\max }\limits_{{\mathbf{B}_1, \mathbf{B}_2, \cdots, \mathbf{B}_K}}
     & ~~R_{\rm sum}^{ {\rm w}} (\mathbf{B}_{1}, \mathbf{B}_{2}, \cdots, \mathbf{B}_{K}), \\
    \text{s.t.} \qquad &  ~~\text{tr} \left(\mathbf{B}_{k}\mathbf{B}^H_{k}\right) \leq P_{k}, ~ \forall k
\end{aligned}
\end{equation}
where
\begin{equation}\label{eq:obj_wsr_3}
 R_{\rm sum}^{ {\rm w}} (\mathbf{B}_{1}, \mathbf{B}_{2}, \cdots, \mathbf{B}_{K}) = \sum\limits_{k = 1}^{K} \Delta_{k} f (\mathbf{B}_{1}, \mathbf{B}_{2}, \cdots, \mathbf{B}_{k})
\end{equation}
with } $\Delta_{k} = \mu _k - \mu _{k+1}$ and $f (\mathbf{B}_{1}, \mathbf{B}_{2}, \cdots, \mathbf{B}_{k}) = I(\mathbf{d}_1, \cdots, \mathbf{d}_k ;
\mathbf{y}|\mathbf{d}_{k+1},\cdots,\mathbf{d}_{K})$, $k = 1,2,\cdots,K$. By evaluating $I(\mathbf{d}_1,\cdots, \mathbf{d}_k ;
\mathbf{y}|\mathbf{d}_{k+1},\cdots,\mathbf{d}_{K})$ based on Proposition \ref{prop:ach_rate_mac}, we obtain an asymptotic expression $R_{\rm sum,asy}^{{\rm w}}
(\mathbf{B}_{1}, \mathbf{B}_{2}, \cdots, \mathbf{B}_{K})$ for $R_{\rm sum}^{{\rm w}} (\mathbf{B}_{1}, \mathbf{B}_{2}, \cdots, \mathbf{B}_{K})$ in
(\ref{eq:obj_wsr_2}). When $\mu_{1} = \mu_{2} = \cdots = \mu_{K} = 1$, (\ref{eq:obj_wsr_2}) reduces to the sum rate maximization.

}

\subsection{Asymptotically Optimal Precoder Structure}
Consider the singular value decomposition (SVD) of the precoder of user $l$,
$\mathbf{B}_l = \mathbf{U}_{{\rm B}_l} \mathbf{\Gamma}_{{\rm B}_l} \mathbf{V}_{{\rm B}_l}$,
where $\mathbf{U}_{{\rm B}_l} $ and $\mathbf{V}_{{\rm B}_l}$ are unitary matrices, and $\mathbf{\Gamma}_{{\rm B}_l}$
is a diagonal matrix with non-negative main diagonal elements. Then, we have the following theorem.
{\bfbl
\begin{theorem}\label{theo:opt_structure}
The left singular matrices $\mathbf{U}_{{\rm B}_l}$ of the asymptotically optimal precoders which maximize the asymptotic WSR $R_{\rm sum,asy}^{{\rm w}}
\left({\mathbf{B}}_1 ,{\mathbf{B}}_2 , \cdots, {\mathbf{B}}_K \right)$  are the eigenmatrices $\mathbf{U}_{{\rm T}_l}$ of the transmit correlation matrices in (\ref{correlation}),
$l = 1,2,\cdots,K$.
Using the new notations below Remark 3 and based on the optimal precoder structure, (\ref{eq:obj_wsr_2}) simplifies to
% Based on \cite[Eq. (8)]{Xiao2011TSP}, the problem (\ref{eq:subproblem}) in Appendix \ref{sec:proof_opt_structure} can be simplified  to
\begin{equation} \label{eq:subproblem_sim}
 \max_{\mathbf{\Gamma}_{{\rm B}_l}, \mathbf{V}_{{\rm B}_l}, \atop{\text{tr} \left(\mathbf{\Gamma}_{{\rm B}_l}^2 \right) \leq P_{l}} }
 ~ \sum\limits_{k = l}^K {{\Delta _k}I\left( {{{\bf{d}}_l};{\bf{z}}_l^{\left( k \right)}\left| {\sqrt {{\rm diag}\left({{\bf{G}}_l^T {\boldsymbol{\gamma}}_l^{\left( k \right)}}\right)} \mathbf{\Gamma}_{{\rm B}_l} \mathbf{V}_{{\rm B}_l}  } \right.} \right)}
     % &~ \text{tr} \left(\mathbf{\Gamma}_{{\rm B}_l}^2 \right) \leq P_{l}, \label{eq:subproblem_constraint_sim}
\end{equation}
for $l = 1,2,\cdots,K$.
\begin{proof}
See Appendix \ref{sec:proof_opt_structure}.
\end{proof}
\end{theorem}
}
\textit{Remark 4:} We note that for finite alphabet input scenarios,  the optimal precoder structure of the left singular matrix has been obtained for point-to-point
MIMO systems \cite{Xiao2011TSP,zeng2012linear}.  {\bfbl Also, the optimal precoder structure for the MIMO MAC was implicitly used in \cite{Girnyk2014TWC}
for the Kronecker model and finite alphabet inputs without proof.  Therefore, the main contribution of Theorem \ref{theo:opt_structure} is the explicit presentation of the optimal precoder structure
for the MIMO MAC for Weichselberger's model and finite alphabet inputs and its proof.}

\textit{Remark 5:} For Weichselberger's model in (\ref{H_channel}), if only sum rate maximization is considered, i.e., $\Delta_1 = \Delta_2$ $ =\cdots = \Delta_{K-1}$, we can directly optimize
 matrix $\mathbf{V}_{{\rm B}_L}^H \mathbf{\Gamma}_{{\rm B}_L}^H {{\rm diag}\left({{\bf{G}}_L^T {\bl \boldsymbol{\gamma}}_L^{\left( k \right)}}\right)} \mathbf{\Gamma}_{{\rm
B}_L} \mathbf{V}_{{\rm B}_L}$  \cite{Girnyk2014TWC} since the mutual information expression in (\ref{eq:subproblem_sim}) is concave
with respect to this matrix.
However, for WSR optimization, since the values of ${\bl \boldsymbol{\gamma}}_l^{\left( k \right)}$ are different for different $k$, it is not possible to find a common matrix
$\mathbf{V}_{{\rm B}_l}^H \mathbf{\Gamma}_{{\rm B}_l}^H {{\rm diag}\left({{\bf{G}}_l^T {\bl \boldsymbol{\gamma}}_l^{\left( k \right)}}\right)} \mathbf{\Gamma}_{{\rm
B}_l} \mathbf{V}_{{\rm B}_l}$ to be optimized in (\ref{eq:subproblem_sim}). {\bl Thus, we optimize  $\mathbf{\Gamma}_{{\rm B}_l}$ and
$\mathbf{V}_{{\rm B}_l}$ in an alternating manner.

\begin{alg} \label{Gradient_MAC}
Iterative  algorithm for  WSR
maximization with respect to $\left\{{\mathbf{B}}_1,{\mathbf{B}}_2 ,\cdots, {\mathbf{B}}_K \right\}$
\vspace*{1.5mm} \hrule \vspace*{1mm}
  \begin{enumerate}

\itemsep=0pt

\item Initialize $\mathbf{\Gamma}_{{\rm B}_l}^{(1)}$, $\mathbf{V}_{{\rm B}_l}^{(1)}$,  $l= 1,2,\cdots,K$, ${\bf{T}}^{(k)}_{t}$, ${\bf{R}}^{(k)}_{t}$,
${\boldsymbol{\psi }}_{t}^{(k), \left( {{1}} \right)}$, and ${\boldsymbol{\gamma }}_{t}^{(k), \left( {{1}} \right)}$, $t = 1,2,\cdots,k$,
 $k = 1,2,\cdots,K$. Set $n = 1$ and compute $R_{\rm sum, asy} ^ {{{\rm w}}, (n)}$.

\item Update $\left({\mathbf{\Gamma}}_{{\rm B}_l}^{(n + 1)}\right)^2$, $l= 1,2,\cdots,K$, along
the gradient decent direction in (\ref{eq:grad_diag}).

\item Update  $\mathbf{V}_{{\rm B}_l}^{(n + 1)}$, $l= 1,2,\cdots,K$, along the gradient decent direction in (\ref{eq:grad_V}).

\item Compute ${\mathbf{B}}_{l}^{(n + 1)} = \mathbf{U}_{{\rm T}_l} \mathbf{\Gamma}_{{\rm B}_l}^{(n + 1)} \mathbf{V}_{{\rm B}_l}^{(n+1)}$ and update
the asymptotic parameters  ${\bf{T}}^{(k)}_{t}$, ${\bf{R}}^{(k)}_{t}$, ${\boldsymbol{\gamma }}_{t}^{(k),\left( {{n + 1}} \right)}$,
and ${\boldsymbol{\psi }}_{t}^{(k),\left( {{n  + 1}} \right)}$.

\item Compute $R_{\rm sum, asy} ^ {{{\rm w}}, (n + 1)}$. If $R_{\rm sum, asy} ^ {{ {\rm w}}, (n + 1)}  - R_{\rm sum, asy} ^ {{{\rm w}}, (n)}$
is larger than a threshold and $n$ is less than the maximal number of iterations,  set $n : = n + 1$, repeat Steps $2$--$4$;
otherwise, stop the algorithm.

 \vspace*{1mm} \hrule

  \end{enumerate}

\end{alg}
\null
\par

Next, we obtain the gradients of $R_{\rm sum,asy}^{\bl {\rm w}}\left({\mathbf{B}}_1
,{\mathbf{B}}_2 , \right.$ $\left.\cdots, {\mathbf{B}}_K \right)$ with respect to $\mathbf{\Gamma}_{{\rm B}_l}^2$ and $\mathbf{V}_{{\rm B}_l}$,
which are given by \cite[Eq. (19)]{Xiao2011TSP} and \cite[Eq. (22)]{Palomar2006TIT} as
\begin{multline}
{\nabla _{{\bf{\Gamma }}_{{{\rm{B}}_l}}^2}}R_{{\rm{sum,asy}}}^{\rm{w}}\left( {{{\bf{B}}_1},{{\bf{B}}_2}, \cdots ,{{\bf{B}}_K}} \right) \\
= \sum\limits_{k = l}^K {{\Delta _k}{\rm{diag}}\left( {{\bf{V}}_{{{\rm{B}}_l}}^H{\bf{E}}_l^{\left( k \right)}{{\bf{V}}_{{{\rm{B}}_l}}}{\rm{diag}}\left(
{{\bf{G}}_l^T{\bl \boldsymbol{\gamma}}_l^{\left( k \right)}} \right)} \right)} \label{eq:grad_diag}
\end{multline}
and
\begin{multline}
{\nabla _{{{\bf{V}}_{{{\rm{B}}_l}}}}}R_{{\rm{sum,asy}}}^{\rm{w}}\left( {{{\bf{B}}_1},{{\bf{B}}_2}, \cdots ,{{\bf{B}}_K}} \right) \\
 = \sum\limits_{k = l}^K {{\Delta _k}{\rm{diag}}\left( {{\bf{G}}_l^T{\bl \boldsymbol{\gamma}}_l^{\left( k \right)}} \right){\bf{\Gamma }}_{{{\rm{B}}_l}}^2} {{\bf{V}}_{{{\rm{B}}_l}}} {\bf{E}}_l^{\left( k \right)}, \label{eq:grad_V}
\end{multline}
respectively. Now, we are ready to propose an iterative algorithm to determine
the optimal precoders $\mathbf{B}_l$ numerically.

{\bl

\subsection{Iterative Algorithm for Weighted Sum Rate Maximization}
Based on Theorem \ref{theo:opt_structure}, (\ref{eq:grad_diag}), and (\ref{eq:grad_V}),
an efficient iterative algorithm can be formulated to determine the optimal precoders $\mathbf{B}_l$ numerically.
The resulting algorithm is summarized in Algorithm \ref{Gradient_MAC}.

{\bfbl
In Step 2 of Algorithm \ref{Gradient_MAC}, we optimize  $\left(\mathbf{\Gamma}_{{\rm B}_l}^{(n)}\right)^2$
along the gradient descent direction $\left(\widetilde{\mathbf{\Gamma}}_{{\rm B}_l}^{(n)} \right)^2 =$ $\left(\mathbf{\Gamma}_{{\rm B}_l}^{(n)}\right)^2
+ u {\nabla _{{\bf{\Gamma }}_{{{\rm{B}}_l}}^2}}R_{{\rm{sum,asy}}}^{\rm{w}}\left( {{{\bf{B}}_1},{{\bf{B}}_2}, \cdots ,{{\bf{B}}_K}} \right)$,
where ${\nabla _{{\bf{\Gamma }}_{{{\rm{B}}_l}}^2}}R_{{\rm{sum,asy}}}^{\rm{w}}\left( {{{\bf{B}}_1},{{\bf{B}}_2}, \cdots ,{{\bf{B}}_K}} \right)$
is given by (\ref{eq:grad_diag}) and the step size $u$ is determined by the backtracking line search method \cite{Boyd2004}.
Thereby, the values of the backtracking line search
 parameters $\theta$ and $\omega$ are set as $\theta \in (0,0.5)$ and $\omega  \in (0,1)$ \cite{Boyd2004}.
If the updated $\left(\widetilde{\mathbf{\Gamma}}_{{\rm B}_l}^{(n)} \right)^2$ has negative elements, we set those to zero,
normalize $\left(\widetilde{\mathbf{\Gamma}}_{{\rm B}_l}^{(n)} \right)^2$ \cite{Girnyk2014TWC}, and then update
$\left({\mathbf{\Gamma}}_{{\rm B}_l}^{(n + 1)} \right)^2 = \left(\widetilde{\mathbf{\Gamma}}_{{\rm B}_l}^{(n)} \right)^2$.
In Step 3, we optimize $\mathbf{V}_{{\rm B}_l}^{(n)}$
along the gradient descent direction $\widetilde{\mathbf{V}}_{{\rm B}_l}^{(n)}   = \mathbf{V}_{{\rm B}_l}^{(n)}
+ u {\nabla _{{\bf{V }}_{{{\rm{B}}_l}}}}R_{{\rm{sum,asy}}}^{\rm{w}}\left( {{{\bf{B}}_1},{{\bf{B}}_2}, \cdots ,{{\bf{B}}_K}} \right)$,
where ${\nabla _{{\bf{V }}_{{{\rm{B}}_l}}}}R_{{\rm{sum,asy}}}^{\rm{w}}\left( {{{\bf{B}}_1},{{\bf{B}}_2}, \cdots ,{{\bf{B}}_K}} \right)$
is given by (\ref{eq:grad_V}). We compute the SVD of $\widetilde{\mathbf{V}}_{{\rm B}_l}^{(n)} = \mathbf{U}_{{\rm{V}}_l} \mathbf{\Gamma}_{{\rm V}_l} \mathbf{V}_{{\rm V}_l}$.
Then, we project $\widetilde{\mathbf{V}}_{{\rm B}_l}^{(n)}$ on the Stiefel manifold
${\mathbf{V}}_{{\rm B}_l}^{(n + 1)} = \mathbf{U}_{{\rm{V}}_l}  \mathbf{V}_{{\rm V}_l} $ \cite[Sec. 7.4.8]{Horn1985}.
In Step 4,  we compute  ${\mathbf{B}}_{l}^{(n + 1)} = \mathbf{U}_{{\rm T}_l} \mathbf{\Gamma}_{{\rm B}_l}^{(n + 1)} \mathbf{V}_{{\rm B}_l}^{(n+1)}$.
Then, we update the
asymptotic parameters  ${\bf{T}}^{(k)}_{t}$, ${\bf{R}}^{(k)}_{t}$, ${\boldsymbol{\gamma }}_{t}^{(k),\left( {{n + 1}} \right)}$,
and ${\boldsymbol{\psi }}_{t}^{(k),\left( {{n  + 1}} \right)}$ in Proposition \ref{prop:ach_rate_mac}
based on the updated precoders ${\bf{B}}_l^{\left( n + 1 \right)}$, $l = 1,2,\cdots,K$, and the fixed point equations (\ref{eq:eqChMatrixTR}) and (\ref{eq:Varsigma_k-MSE}).
In Step 5, we compute $R_{\rm sum, asy} ^ {{{\rm w}}, (n + 1)}$ based on  ${\mathbf{B}}_{l}^{(n + 1)}$, $l = 1,2,\cdots,K$,  ${\bf{T}}^{(k)}_{t}$, ${\bf{R}}^{(k)}_{t}$, ${\boldsymbol{\gamma }}_{t}^{(k),\left( {{n + 1}} \right)}$, and ${\boldsymbol{\psi }}_{t}^{(k),\left( {{n + 1}} \right)}$, $t = 1,2,\cdots,k$, $k = 1,2,\cdots,K$. Finally, if $R_{\rm sum, asy} ^ {{ {\rm w}}, (n + 1)}  - R_{\rm sum, asy} ^ {{{\rm w}}, (n)}$
is larger than a threshold and $n$ is less than the maximal number of iterations, we perform the next iteration, otherwise, we stop the algorithm.
}

{\bl \textit{Remark 6:} We note that the iterative algorithms in \cite{Xiao2011TSP} and \cite{zeng2012linear} are for point-to-point MIMO systems. Therefore, both
algorithms only need to consider the maximization of a {single} mutual information expression. {For the MIMO MAC}, the iterative algorithm in
\cite{Wang2011} optimizes the precoders of all  users jointly. Therefore, its implementation complexity is very high.
Based on the asymptotic WSR expression, Algorithm 1 optimizes the precoder of each user separately. Thus, its implementation complexity is significantly lower than that of the iterative algorithm in \cite{Wang2011}. Moreover, Algorithm 1 exploits the optimal structure of the precoders for WSR maximization
 and optimizes the power allocation matrix and the right singular  matrix of the precoder of each user in an alternating manner.
The iterative algorithm in
\cite{Girnyk2014TWC} is for the Kronecker model.  {\bfbl For the special case of the Kronecker model, even for  WSR optimization, ${\boldsymbol{\gamma}}_l^{\left( k \right)}$ can  be moved
outside $\mathbf{V}_{{\rm B}_l}^H  \mathbf{\Gamma}_{{\rm B}_l}^H {{\rm diag} \left({{\bf{G}}_l^T {\boldsymbol{\gamma}}_l^{\left( k \right)}}\right)} $  $  \mathbf{\Gamma}_{{\rm
B}_l}  \mathbf{V}_{{\rm B}_l}$, as indicated in \cite{Girnyk2014TWC}. Hence, for the Kronecker model,
the algorithm in \cite{Girnyk2014TWC}  can also be used to optimize the WSR and may be
 preferable for numerical calculation. This is because the algorithm in \cite{Girnyk2014TWC} only requires
 an eigenvalue decomposition where Algorithm 1 requires a SVD. However,  as indicated in Remark 5, the algorithm in \cite{Girnyk2014TWC} can not be directly applied to the WSR optimization problem for Weichselberger's model
considered in this paper. }

{\bl \textit{Remark 7:}
We note that calculating the mutual information and the MSE matrix
(e.g., (\ref{eq:mutual_info}), (\ref{eq:grad_diag}), (\ref{eq:grad_V}) or \cite[Eq. (5)]{Wang2011}, \cite[Eq. (24)]{Wang2011}) involves additions over the modulation signal space
which scales exponentially with the number of transmit antennas. The computational complexity of other operations, such as
the matrix product,  solving the fixed point equations, etc., are polynomial functions of the number of transmit and receive antennas.
Therefore, for ease of analysis, we just compare the computational complexity of
calculating the mutual information and the MSE matrix here. When $N_t$ increases, the computational complexity of Algorithm \ref{Gradient_MAC} is dominated by the required number of additions in calculating
$R_{\rm sum,asy} ^ {{\bl {\rm w}},(n)} \left({\mathbf{B}}_1 ,{\mathbf{B}}_2 , \cdots, {\mathbf{B}}_K \right)$,
${\nabla _{{\bf{\Gamma }}_{{{\rm{B}}_l}}^2}}R_{{\rm{sum,asy}}}^{\rm{w}}\left( {{{\bf{B}}_1},{{\bf{B}}_2}, \cdots ,{{\bf{B}}_K}} \right)$, and ${\nabla _{{\bf{V }}_{{{\rm{B}}_l}}}}R_{{\rm{sum,asy}}}^{\rm{w}}$
$\left( {{{\bf{B}}_1},{{\bf{B}}_2} },\cdots, {{{\bf{B}}_K}} \right)$ based on (\ref{eq:mutual_info})
in Steps 1 and 5, Step 2, and Step 3, respectively. Eq. (\ref{eq:mutual_info}) implies that Algorithm \ref{Gradient_MAC} only requires additions over each user's own
possible transmit vectors to design the precoders.  Accordingly, the  computational complexity\footnote{\bl
 The average over  the noise vector $\mathbf{v}$ in (\ref{eq:mutual_info}) can be evaluated by employing the
  accurate approximation in \cite[Eq. (57)]{zeng2012linear}. Therefore, its
computational burden is negligible compared to that of computing the expectation over $\mathbf{d}_{i_k}$ in (\ref{eq:mutual_info}).}
of the proposed Algorithm \ref{Gradient_MAC} in calculating the mutual
information and the MSE matrix  grows linearly with $\sum\nolimits_{k = 1}^K {Q_k^{2 N_t } }$.
In contrast, the conventional precoder
design for instantaneous CSI at the transmitter in \cite{Wang2011} requires additions over all possible transmit
vectors of all users. For this reason, the computational complexity of the conventional precoding
design  scales linearly with $\left( {\prod\nolimits_{k = 1}^K {Q_k } } \right)^{2 N_t }$.
As a result, the computational complexity of Algorithm \ref{Gradient_MAC} is significantly lower than that of the conventional design.
We note that this computational complexity reduction is more obvious when the number of transmit antennas or the number of users become large. }

To show this more clearly,  we give an example. We consider a practical massive MIMO MAC system where the base station is equipped with a large number of antennas and
serves multiple users having much smaller numbers of antennas \cite{Marzetta2010TWC,Rusek2013SPM,Adhikary2013TIT,Larsson2013,Liang2014,Wen2014}.
 In particular, we assume $N_r = 64$, $N_t = 4$, $K = 4$,
$\mu_1 = \mu_2 = \mu_3 = \mu_4$, and all users
employ the same modulation constellation.  The numbers  of additions required for calculating the mutual information and the MSE matrix
in Algorithm \ref{Gradient_MAC} and in the precoder
design in \cite{Wang2011} are listed in Table \ref{tab:mac_dim} for different modulation formats.
}

\begin{table}[!t]
\centering
% \captionstyle{center}
\caption{Number of additions required for calculating the mutual information and the MSE matrix.} \label{tab:mac_dim}
\vspace*{1.5mm}
\begin{tabular}{|c|c|c|c|c|c|c|}
\hline
\  Modulation     &  QPSK & 8PSK &   16 QAM     \\ \hline
\  Algorithm \ref{Gradient_MAC}       &   262144  & 6.7 e+007  &  1.7 e+010    \\ \hline
 \  Design Method in \cite{Wang2011} & 1.85  e+019  & 7.9 e+028  &  3.4 e+038       \\ \hline
\end{tabular}
\end{table}

We observe from Table \ref{tab:mac_dim} that Algorithm \ref{Gradient_MAC} requires a significantly
lower number of  additions  for  the MIMO MAC precoder design for finite alphabet inputs compared to the design
in \cite{Wang2011}.
Moreover, since Algorithm \ref{Gradient_MAC} is based on the channel statistics
$\{\mathbf{U}_{{\rm T}_k}\}_{\forall k}$, $\{\mathbf{U}_{{\rm R}_k}\}_{\forall k}$, $\{\mathbf{G}_k\}_{\forall k}$,
it avoids the time-consuming averaging process over each channel realization of
the mutual information in (\ref{mutual_info_1}). In addition, Algorithm \ref{Gradient_MAC} is executed only once since
the precoders are constant as long as
the channel statistics do not change, whereas the algorithm in \cite{Wang2011} has to be executed for each channel realization.

{\bl \textit{Remark 8:} We note that Algorithm \ref{Gradient_MAC}  never decreases
the asymptotic WSR $R_{\rm sum,asy} ^ {\rm w} \left({\mathbf{B}}_1 ,{\mathbf{B}}_2 , \cdots, {\mathbf{B}}_K \right)$ in any iteration, see Step 5.
From the expression in (\ref{eq:GAUMutuall_2}), we also know that the asymptotic WSR $R_{\rm sum,asy} ^ {\rm w} \left({\mathbf{B}}_1 ,{\mathbf{B}}_2 , \cdots, {\mathbf{B}}_K \right)$
is upper-bounded. This implies that
Algorithm \ref{Gradient_MAC}, which produces non-decreasing sequences that are upper-bounded, is convergent. Due to the non-convexity of the objective function $R_{\rm sum,asy}^{\bl {\rm w}} \left({\mathbf{B}}_1 ,{\mathbf{B}}_2
, \cdots, {\mathbf{B}}_K \right)$, in general, Algorithm \ref{Gradient_MAC} will find a local maximum of the WSR. Therefore, we run Algorithm 1 for several random
initializations ${\mathbf{B}}_{k}^{(1)}$ and select the result that offers the maximal WSR as the final design solution \cite{Wang2011,Perez-Cruz2010TIT}. }

\section{Numerical Results}
In this section, we provide examples to illustrate the performance
of the proposed iterative optimization algorithm.
We assume equal individual power limits $P_1 = P_2 =\cdots =P_{K} = P$
and the same modulation format for all $K$ users.
The average SNR for the MIMO MAC with statistical CSI is defined as
${\rm{SNR}} = \frac{{E\left[ {{{\rm{tr}}\left( {{\bf{H}}_k {\bf{H}}_k^H } \right)}} \right] P}}{{N_t N_r }}$.
{\bl We use GP, NP, FAP, and AL as abbreviations for Gaussian precoding, no precoding, finite alphabet precoding,
and algorithm in \cite{Wang2011}, respectively.}

{\bl First, we consider a two-user
MIMO MAC  with two transmit antennas for each user and two
receive antennas
in order to illustrate that,
although Algorithm \ref{Gradient_MAC} was derived for the large system limit,
it also performs well if the numbers of antennas are small.}
{\bfbl For the channel statistics of Weichselberger's model,
 ${\bf{U}}_{{\rm{T}}_k}$, ${\bf{U}}_{{\rm{R}}_k }$, and
${{ \bf\widetilde{G}}_k}$, $k = 1,2$, are chosen at random. }

Figure \ref{Exact_Asy} depicts the average exact sum rate obtained based on (\ref{eq:Finite_Mutual})
and the sum rate obtained with the asymptotic expression in Proposition
\ref{prop:ach_rate_mac} for different precoding designs and QPSK inputs.
For the case without precoding, we set ${\mathbf{B}}_1
 = {\mathbf{B}}_2  = \sqrt{\frac{P}{N_t}} \mathbf{I}_{N_t}$,
{\bl and denote the corresponding exact and asymptotic sum rates as ``NP, Exact"
 and ``NP, Asymptotic", respectively.
Furthermore, we denote the exact and asymptotic sum rates achieved by the design proposed in Algorithm 1 as ``FAP, Exact" and ``FAP, Asympotic".
From Figure \ref{Exact_Asy}, we observe
that the asymptotic sum rate expression in
Proposition \ref{prop:ach_rate_mac} provides a good estimate of the exact sum rate even for small numbers of
antennas. } On the other hand, if the numbers of antennas are large,
evaluating the exact mutual information in (\ref{mutual_info_1}) numerically via Monte Carlo simulation is extremely
time-consuming. {\bl In contrast}, Proposition \ref{prop:ach_rate_mac} provides an efficient method for estimating the
ergodic WSR of the MIMO MAC  with finite alphabet inputs.

\begin{figure}[!t]
\centering
\includegraphics[width=0.4\textwidth]{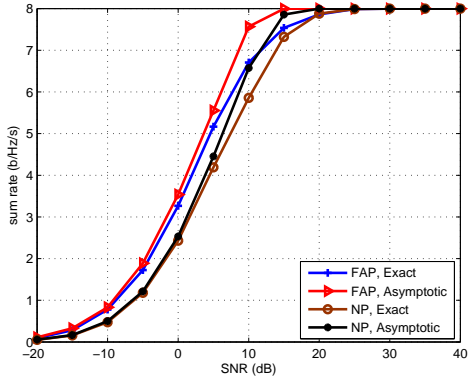}
\caption {\space\space Average sum rate for different precoding designs. }
\label{Exact_Asy}
\end{figure}

Figure \ref{Convergence} illustrates the convergence behavior of  Algorithm 1 for different SNR values and QPSK inputs.
{\bl We set the backtracking line search  parameters to $\theta  = 0.1$ and $\omega = 0.5$.} Figure \ref{Convergence} shows
the sum rate in each iteration. We observe
that in all considered cases, the proposed algorithm needs only
a few iterations to converge.

\begin{figure}[!t]
\centering
\includegraphics[width=0.4\textwidth]{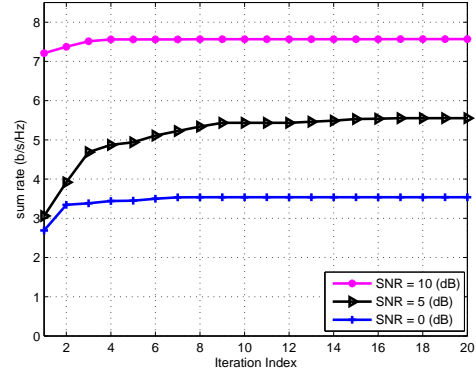}
\caption {\space\space Sum rate vs. iteration index for different SNRs, QPSK inputs,  $\theta = 0.1$, and $\omega = 0.5$.}
\label{Convergence}
\end{figure}

In Figure \ref{Sum_Rate_QPSK_mac}, we show the sum rate for different transmission schemes and
QPSK inputs. We employ the Gauss-Seidel algorithm together with stochastic programming\footnote{\bl{
We note that the asymptotically optimal precoder design for Gaussian input signals
in \cite{Romain2011TIT} has a concise structure and a low implementation complexity.
However, the main purpose of considering the precoder design under the Gaussian input assumption in this paper
is to show that
 the Gaussian input assumption
precoder design departs remarkably from the practical finite alphabet input design.
Therefore, we consider the Gauss-Seidel algorithm together with stochastic programming for precoder design as this method
 optimizes the exact sum rate of the MIMO MAC with Gaussian input.
Although this approach is complicated, it achieves the best sum rate performance for Gaussian input.}}
to obtain the optimal covariance matrices of the users under the Gaussian input assumption \cite{wen2011sum}.
Then, we decompose the obtained
optimal covariance matrices $\left\{ {{\mathbf{Q}}_1 , {\mathbf{Q}}_2, \cdots ,{\mathbf{Q}}_K } \right\}$
as ${\mathbf{Q}}_k = {\mathbf{U}}_k \boldsymbol{\Lambda} _k {\mathbf{U}}_k^H$, and set ${\mathbf{B}}_k
= {\mathbf{U}}_k \boldsymbol{\Lambda} _k^{\frac{1}{2}}$, $k = 1,2, \cdots, K$. Finally, we calculate
the average sum rate for this precoding design for QPSK inputs. We denote the corresponding sum rate
as ``GP with QPSK inputs".
For the case without precoding, we set ${\mathbf{B}}_1
 = {\mathbf{B}}_2  = \sqrt{\frac{P}{N_t}} \mathbf{I}_{N_t}$.
 We denote the corresponding sum rate as ``NP with QPSK inputs".
 We denote the proposed design in Algorithm 1 as ``FAP with QPSK inputs".
 For comparison purpose, we also show the average sum rate achieved by Algorithm 1 in \cite{Wang2011}
 with instantaneous CSI and denote it as ``AL in [12] with QPSK inputs".
The sum rates achieved with the Gauss-Seidel algorithm and without precoding for Gaussian inputs
 are also plotted in Figure \ref{Sum_Rate_QPSK_mac}, and are denoted as ``GP with Gaussian input" and
``NP with Gaussian input", respectively.

\begin{figure}[!t]
\centering
\includegraphics[width=0.4\textwidth]{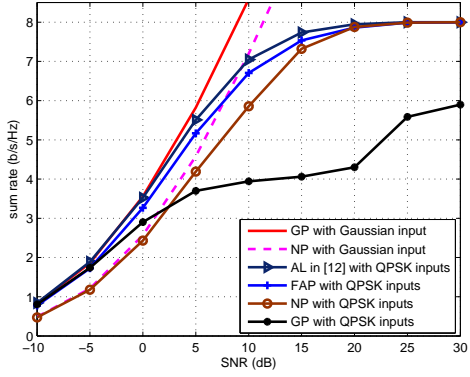}
\caption {\space\space Average sum rate of two-user MIMO MAC with QPSK modulation.}
\label{Sum_Rate_QPSK_mac}
\end{figure}

From Figure \ref{Sum_Rate_QPSK_mac}, we make the following observations: 1) For QPSK modulation, the proposed iterative algorithm
achieves a considerably higher
sum rate compared to the other statistical CSI based precoder designs.
Specifically, to achieve a target sum rate of $4$ b/s/Hz, the proposed algorithm achieves SNR gains of approximately
 $2.5$ dB and $11$ dB compared to the ``NP with QPSK inputs" design and
the ``GP with QPSK inputs" design, respectively.
2) The sum rate achieved by the proposed algorithm is
close to the sum rate achieved by Algorithm 1 in \cite{Wang2011} which requires instantaneous CSI.
At a target sum rate of $4$ b/s/Hz, the SNR gap between the proposed algorithm and
Algorithm 1 in \cite{Wang2011} is less than 1 dB.  However, the proposed algorithm only requires
statistical CSI and its implementation complexity is much lower than that of Algorithm 1 in \cite{Wang2011}.
3) The sum rate achieved by the proposed algorithm and the ``NP with QPSK inputs"
design merge at high SNR,  and both saturate at $K\log_2 M = 8$ b/s/Hz.
4) The sum rate achieved by the ``GP with QPSK inputs" design  remains almost constant
for SNRs between $10$ dB and $20$ dB. This is because the Gauss-Seidel algorithm design implements
a ``water filling" power allocation policy in this SNR region.
As a result, when the SNR is smaller than a threshold (e.g., 20 dB in this case), the precoders allocate most of the available
power to the strongest subchannels and allocate little power to the weaker subchannels. Therefore,
one eigenvalue of ${\mathbf{Q}}_k$  approaches zero.
For example, for $\rm{SNR} = 10$ dB, the optimal
covariance matrices obtained by the Gauss-Seidel algorithm are given by
\begin{eqnarray}\label{Q_1}
{\mathbf{Q}}_1 & = & \left[ \begin{array}{l}
   1.2599     \qquad    \qquad  \quad \ 0.9619 + 0.0790j \\
   0.9619 - 0.0790j  \quad  0.7401   \\
 \end{array} \right] \nonumber \\
 {\mathbf{Q}}_2  & = & \left[ \begin{array}{l}
  \ \ \, 0.1495        \qquad \qquad \qquad  -0.5214 + 0.0651j \\
  -0.5214 - 0.0651j \qquad \ \   1.8505    \\
 \end{array} \right]. \nonumber \\
\end{eqnarray}
After  eigenvalue decomposition
${\mathbf{Q}}_k = {\mathbf{U}}_k \boldsymbol{\Lambda} _k {\mathbf{U}}_k^H$, we have
\begin{eqnarray}\label{lambda}
\boldsymbol{\Lambda} _1 & = & \rm{diag}\left\{9.9976,0.0024 \right\}, \nonumber \\
\boldsymbol{\Lambda} _2 & = & \rm{diag}\left\{9.9987,0.0013\right\} .
\end{eqnarray}
The precoders are given by
\begin{eqnarray}\label{G_g}
{\mathbf{B}}_1  = \left[ \begin{array}{l}
  -2.5097 - 0.0000j  \qquad  0.0298 - 0.0000j \\
  -1.9168 + 0.1574j \quad -0.0388 + 0.0032j  \\
 \end{array} \right] \nonumber \\
 {\mathbf{B}}_2  = \left[ \begin{array}{l}
  -0.8638 + 0.0000j \quad -0.0349 - 0.0000j \\
 \ \  3.0184 + 0.3767j \quad \ -0.0098 - 0.0012j \\
 \end{array} \right].
\end{eqnarray}

From the structure of the precoders in (\ref{G_g}),
we can see that most energy is allocated  to one transmitted symbol.
For finite alphabet inputs, this power allocation policy may result in allocating most power to the subchannels
that are close to saturation. This will lead to a waste of transmit power and impede the
further improvement of the sum rate performance. This
confirms that precoders designed under the ideal Gaussian input assumption may  result in
a considerable performance loss when adopted directly in practical systems with finite alphabet
constraints.

\begin{figure}[!t]
\centering
\includegraphics[width=0.4\textwidth]{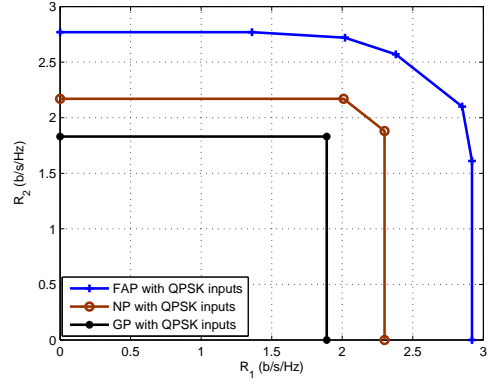}
\caption {\space\space Achivable rate regions of two-user MIMO MAC with QPSK modulation.}
\label{rate_region_qpsk}
\end{figure}

{\bl In Figure \ref{rate_region_qpsk}, we show the achievable rate region of different precoder designs
for ${\rm SNR} = 5$ dB and QPSK inputs. The achievable rate regions are
obtained by solving the WSR optimization problem in (\ref{eq:obj_wsr_2}) for different precoder designs.
 We observe from Figure 4 that the proposed design has a much larger rate region
 than the case without precoding and the design based on the Gaussian input assumption.
 We note that since for GP,  most energy is allocated to one transmitted symbol,
  for finite alphabet inputs, the achievable sum rate of this transmission design may result in
 a value that is even smaller than the single user rate. Therefore, there is only one point in the achievable rate region
 for the GP design. A similar phenomenon has also been observed for the MIMO MAC with instantaneous CSI and finite alphabet inputs,
 see \cite[Fig. 6]{Wang2011}. }

To further validate the performance of the proposed design, Figure \ref{Sum_Rate_16QAM_mac} shows the sum rate performance
for different precoding schemes
for 16QAM modulation.
Figure \ref{Sum_Rate_16QAM_mac} indicates that the proposed design outperforms
the other precoding schemes\footnote{We note that
the implementation complexity of Algorithm 1 in \cite{Wang2011} is prohibitive for 16QAM inputs. In contrast,
the complexity of the algorithm proposed in this paper is manageable for 16QAM inputs throughout the entire SNR region.} also for 16QAM modulation. At a sum rate of $8$ b/s/Hz,
the proposed algorithm achieves SNR gains of about $1.7$ dB and $7.5$ dB over
the ``NP with 16QAM inputs" design
and the ``GP with 16QAM inputs" design, respectively.

\begin{figure}[!t]
\centering
\includegraphics[width=0.4\textwidth]{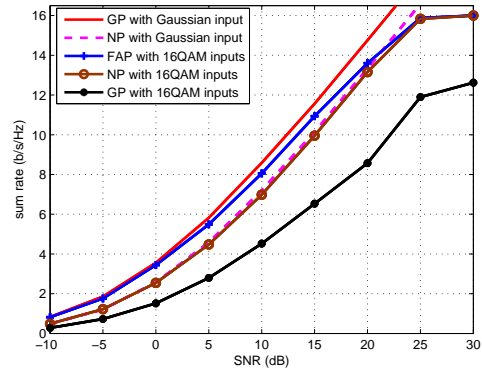}
\caption {\space\space Average sum rate of two-user MIMO MAC with 16QAM modulation.}
\label{Sum_Rate_16QAM_mac}
\end{figure}

\begin{figure}[!t]
\centering
\includegraphics[width=0.4\textwidth]{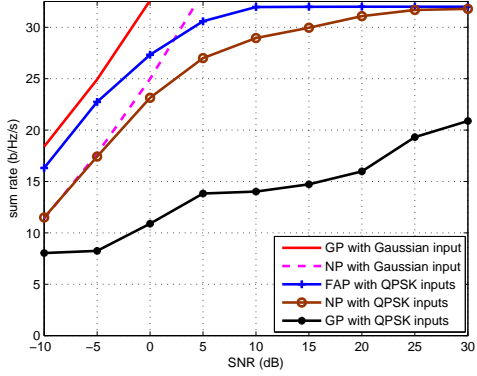}
\caption {\space\space Average sum rate of four-user massive MIMO MAC in suburban scenario with QPSK modulation. }
\label{sum_rate_qpsk_64_suburban}
\end{figure}

In the following, we investigate the performance of the proposed precoder design in a practical massive MIMO MAC system where the base station
is equipped with a large number of antennas and simultaneously serves multiple users
with much smaller numbers of antennas \cite{Marzetta2010TWC,Rusek2013SPM,Adhikary2013TIT,Larsson2013}.
We assume $N_r = 64$ and $N_t = 4$. Furthermore, we adopt the 3rd generation partnership project
spatial channel model (SCM) in \cite{Salo2005}.  We set\footnote{The SCM simulation model in \cite{Salo2005} has several system parameters,
including the number of user, the numbers of antenna, the antenna spacing, the velocity of the users, etc. After setting these parameters, we generated
a large number of channel realizations and calculated the statistical
CSI based on these channel realizations.} the transmit and receive antenna spacings
to half a wave length,
and the velocity\footnote{We consider the scenario where the mobility of the users is high. In such a scenario, it is reasonable to exploit the statistical CSI at the transmitter for precoder design \cite{Gao}.} of the users to $180$ km/h. Figures \ref{sum_rate_qpsk_64_suburban}
and \ref{sum_rate_qpsk_64_urban} show the sum rate performance for different precoder designs, $K=4$, and QPSK inputs
for the suburban and the urban scenarios of the SCM, respectively. We observe from  Figures \ref{sum_rate_qpsk_64_suburban}
and \ref{sum_rate_qpsk_64_urban} that, for QPSK inputs,
the proposed algorithm achieves a better performance than the other precoder designs for
both scenarios.  For a sum rate of $24$ b/s/Hz,
the SNR gains of the proposed algorithm over the ``NP with QPSK inputs" design for the
suburban and the urban scenarios are about $5$ dB and $4.5$ dB, respectively.
The SNR gain for the suburban scenarios is larger than that for the urban scenarios, since the correlation of the transmit antennas
is stronger in suburban scenarios. As a result, the precoder design based on  statistical
CSI is more effective and yields a larger performance gain.
Also, the ``GP with QPSK inputs" design results in a substantial performance loss in both scenarios.
{\bl To illustrate the importance of designing the precoders for  Weichsenberger's channel model, we also show in Figure \ref{sum_rate_qpsk_64_urban}
the sum rate performance of  precoders designed for the Kronecker's model (\emph{i.e.}, the mutual coupling is ignored for precoder calculation)
and denote the corresponding curve as ``FAP with QPSK inputs and KR". We observe from Figure \ref{sum_rate_qpsk_64_urban} that for a sum rate of $24$ b/s/Hz,
we lose about $1$ dB in performance if we design the precoders for the Kronecker's model. }

Figure \ref{Sum_Rate_qpsk_64_urban_user} shows the average sum rate
for different precoder designs as a function of
the number of users for  QPSK inputs,
the urban scenario, and $\rm{SNR} = 0$ dB. We observe
from Figure \ref{Sum_Rate_qpsk_64_urban_user} that the average sum rate
scales linearly with the number of users. This coincides with the conclusion
in Proposition \ref{prop:ach_rate_mac}
that the sum rate can be approximated by the sum of the
individual rates of all users.

\begin{figure}[!t]
\centering
\includegraphics[width=0.4\textwidth]{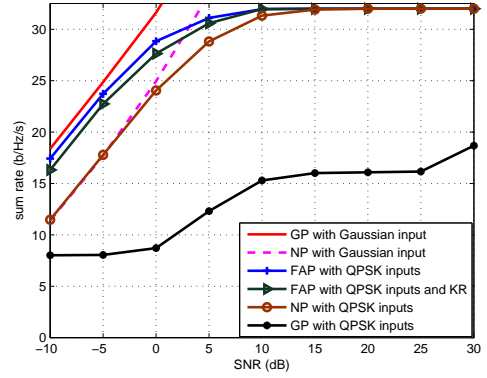}
\caption {\space\space Average sum rate of four-user massive MIMO MAC in urban scenario with QPSK modulation.}
\label{sum_rate_qpsk_64_urban}
\end{figure}

\begin{figure}[!t]
\centering
\includegraphics[width=0.4\textwidth]{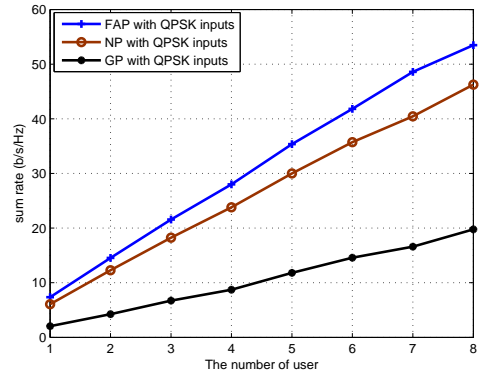}
\caption {\space\space Average sum rate of massive MIMO MAC in urban scenario with QPSK modulation.}
\label{Sum_Rate_qpsk_64_urban_user}
\end{figure}

\section{Conclusion}
In this paper, we have studied the linear precoder design
for the $K$-user MIMO MAC with statistical CSI at the transmitter.
We formulated the problem from the standpoint of finite alphabet inputs based on Weichselberger's MIMO channel model.
We first obtained the WSR expression for the MIMO MAC assuming  Weichselberger's model for the asymptotic large system regime
under the finite alphabet input constraint.
{\bl Then, we established the optimal structures of the precoding matrices which maximize the
asymptotic WSR. } Subsequently, we proposed an iterative
algorithm to find the precoding matrices of all users for  statistical CSI at the transmitter.
We show that the proposed algorithm significantly reduces the implementation complexity
compared to a previously proposed precoder design method for the
MIMO MAC with finite alphabet inputs and instantaneous CSI at the transmitter.
Numerical results showed that, for finite alphabet inputs, precoders designed with
the proposed iterative algorithm achieve substantial performance gains
over  precoders designed based on the Gaussian input assumption and transmission without precoding.
These gains can be observed for both MIMO  systems with
small numbers of antennas and massive MIMO systems.

\appendices
\section{Proof of Proposition \ref{prop:ach_rate_mac}}\label{sec:proof_ach_rate_mac}
Before we present the proof, we introduce the following three useful lemmas.
\begin{lemma} \label{Lemma_2}
Let $\mathbf{S} \in \mathbb{C}^{m \times n}$, $\mathbf{A}_1 \in \mathbb{C}^{m \times n}$, and $\mathbf{A}_2 \in \mathbb{C}^{m \times n}$  be  complex matrices and $\mathbf{A}_3 \in \mathbb{C}^{n \times n} $ and $\mathbf{A}_4 \in \mathbb{C}^{m \times m} $ positive definite  matrices, respectively. Then, the
following equality holds \cite{Mou-03}:
\begin{multline}
\int  D \mathbf{S}  \, e^{-{\tr}\left(\mathbf{A}_3 \mathbf{S}^H \mathbf{A}_4 \mathbf{S} + \mathbf{A}_1^H \mathbf{S}-\mathbf{S}^H
\mathbf{A}_2\right)} \\
 =\frac{1}{\det(\mathbf{A}_3 \otimes \mathbf{A}_4)}e^{-{\tr}\left(\mathbf{A}_3^{-1} \mathbf{A}_1^H \mathbf{A}_4^{-1} \mathbf{A}_2\right)}.
\end{multline}
\end{lemma}
For $\mathbf{A}_1 = \mathbf{A}_2 = {\bf 0}$ and Gaussian random matrix $\mathbf{S}$, we obtain with this lemma the useful result
\begin{equation}
\int   D \mathbf{S} \, e^{-{\sf tr}\left( \mathbf{A}_3 \mathbf{S}^H \mathbf{A}_4 \mathbf{S} \right)}  = \frac{1}{\det(\mathbf{A}_3 \otimes \mathbf{A}_4)}.
\end{equation}

\begin{lemma} \label{Lemma_3}
The eigen-decomposition of matrix ${\bf A}=b{\bf 11}^H+(a-b)\mathbf{I}_{r + 1}\in {\mathbb C}^{(r+1) \times (r+1)}$, where ${\bf{1}} \in \mathbb{C}^{ (r + 1) \times 1}$ is the
 all-one vector, and $a$ and $b$ are arbitrary constants,  is
\begin{equation}
{\bf A}={\bf F} \ {\rm diag}\left(a + r b, a-b, \cdots, a-b \right) {\bf F}^H
\end{equation}
where $ \mathbf{F} \in {\mathbb C}^{(r+1) \times 1}$ is the discrete Fourier transform matrix with elements $[{\bf F}]_{nm}=\frac{1}{\sqrt{r+1}} e^{-j \frac{2\pi}{r+1}(n-1)(m-1)}$.
\end{lemma}

\begin{lemma} \label{Lemma_4}
Hubbard-Stratonovich Transformation:  Let ${\bf s}$ and ${\bf a}$ be arbitrary $m\times 1$ complex vectors. Then, we have
\begin{equation}
e^{{\bf a}^\dag{\bf a}}=\int D {\bf s}e^{-({\bf s}^\dag{\bf s}-{\bf a}^\dag{\bf s}-{\bf s}^\dag{\bf a})}.
\end{equation}
The identity can be proven easily by using the definition of a matrix variate Gaussian distribution. The transformation is a convenient tool to reduce a quadratic
form to a linear expression by introducing auxiliary variables \cite{Nishimori2001}.
\end{lemma}

Now, we begin with the proof of Proposition \ref{prop:ach_rate_mac}. {\bl We note that throughout this section, the virtual channel model defined in Section III-A
is only used if explicitly stated.} First, we consider the case $K_1 = K$. Define ${\bf{H}} = \left[ {{\bf{H}}_{1} \ {\bf{H}}_{2}  \cdots
\mathbf{H}_{{K}} } \right]$, $\mathbf{B} = {\rm{blockdiag}} \left\{ \mathbf{B}_{1},\mathbf{B}_{2}, \cdots, \right. $ $ \left. \mathbf{B}_{K} \right\}$, $
\mathbf{x} = \left[ \mathbf{x}_{1}^T \ \mathbf{x}_{2}^T  \cdots \mathbf{x}_{K}^T \right]^T$, and $ \mathbf{d} = \left[ \mathbf{d}_{1}^T \ \mathbf{d}_{2}^T  \cdots
\mathbf{d}_{K}^T \right]^T$. From (\ref{eq:Finite_Mutual}), the mutual information of the MIMO MAC can be expressed as $ I(\mathbf{d}; {\bf y})= F - N_r \log_2 e
$, where $ F = -{E}_{{\bf y},\mathbf{H}}\left[\log_2 Z({\bf y},\mathbf{H})\right]$ and $Z({\bf y},\mathbf{H}) = {E}_{\mathbf{x}}\left[e^{- \left\|{\bf y}-
\mathbf{Hx}\right\|^2}\right]$. The expectations over $\mathbf{y}$ and $\mathbf{H}$ are difficult to perform because the logarithm appears inside the average. {\bl
The replica method \cite{Edwards1975} circumvents this difficulty by rewriting $F$ as
\begin{equation} \label{eq:ap_sf_F}
F = -\log_2 e \lim_{r\rightarrow 0}\frac{\partial}{\partial r}\ln{ E}_{{\bf y},\mathbf{H}}\left[\left(Z({\bf y},\mathbf{H})\right)^r\right].
\end{equation}
This reformulation is very useful because it allows us to first evaluate ${E}_{{\bf y},\mathbf{H}}\left[\left(Z({\bf y},\mathbf{H})\right)^r \right]$ for a positive integer-valued $r$,
and then extend the result to $r\rightarrow 0$. Note, however, that the replica method is not rigorous. Nevertheless,
it has been widely adopted in the field of statistical physics \cite{Nishimori2001} and  has been also used to derive a number of interesting results in information and communication
theory \cite{wen2007asymptotic,Tanaka2002TIT,RMuller2008JSAC,Guo2005TIT,wen2011sum,Mou-03,Zaidel2012TIT}. Some results obtained based on the replica method have been recently confirmed by more
rigorous analyses, see e.g. \cite{Hachem2008TIT,Korada2011TIT}.
}

%The reformulation is very useful because it allows us to first evaluate ${E}_{{\bf y},\mathbf{H}}\left[\left(Z({\bf y},\mathbf{H})\right)^r \right]$ for an
%integer-valued $r$, before considering
% $r$ in the vicinity of $0$.

In a first step,  to compute the expectation over $Z({\bf y},\mathbf{H})$, it is useful to introduce $r + 1$ replicated signal vectors $\mathbf{x}_k^{(\alpha)}$, for $\alpha = 0, 1, \cdots,
r$, yielding
\begin{equation} \label{eq:ap_sf_E1}
{ E}_{{\bf y},\mathbf{H}}\left[\left(Z({\bf y},\mathbf{H})\right)^r \right] \! = \!{ E}_{\mathbf{H},\mathbf{X}}\left[\int D \mathbf{y} \prod_{\alpha=0}^re^{- \left\|{\bf y}-\sum_{k=1}^{K} \mathbf{H}_k \mathbf{x}_k^{(\alpha)}\right\|^2}\right]
\end{equation}
where $\mathbf{X} = \left[\mathbf{X}_1^T \, \mathbf{X}_2^T \, \cdots \, \mathbf{X}_K^T \right]^T$, $\mathbf{X}_k = \left[ \mathbf{x}_k^{(0)} \,\mathbf{x}_k^{(1)}\,
\cdots \, \mathbf{x}_k^{(r)} \right]$, and the $\{\mathbf{x}_k^{(\alpha)}\}$ are i.i.d. with distribution $p(\mathbf{x}_k)$. Now, the {\bl integration} over
$\mathbf{y}$ can be performed in (\ref{eq:ap_sf_E1}) because it is reduced to the Gaussian integral. {\bl However, the expectations over $\mathbf{H}$ and $\mathbf{X}$ are involved.} To
tackle this problem, we separate the expectations with respect to $\mathbf{X}$ and $\mathbf{H}$. Towards this end, define a set of random matrices: $\mathbf{V} = \left[
\mathbf{V}_1 \, \mathbf{V}_2 \, \cdots \, \mathbf{V}_K \right]$, $\mathbf{V}_k = \left[ \mathbf{v}_{k,1}^T \, \mathbf{v}_{k,2}^T \,\cdots \mathbf{v}_{k,N_r}^T
\right]^T$ and random vectors $\mathbf{v}_{k,n} = \sum_{m}\mathbf{v}_{k,n,m}$, $\mathbf{v}_{k,n,m} = \left[v_{k,n,m}^{(0)} \, v_{k,n,m}^{(1)} \, \cdots \, v_{k,n,m}^{(r)} \right]$,
and $v_{k,n,m}^{(\alpha)} = [\mathbf{W}_k]_{n,m} [\tilde{\mathbf{G}}_k]_{n,m} \mathbf{u}_{{\rm T}_k,m}^{H}$ $\mathbf{x}_k^{(\alpha)}$ for $\alpha=0,1,\cdots,r$.
Then, we have from (\ref{H_channel})
\begin{equation} \label{eq:HforV}
 \mathbf{H}_k \mathbf{x}_k^{(\alpha)} = \sum_{n=1}^{N_r} \left(\sum_{m=1}^{N_t} v_{k,n,m}^{(\alpha)} \right) \mathbf{u}_{{\rm R}_k,n}.
\end{equation}
Notice that, for given $\mathbf{X}_k$, $\mathbf{v}_{k,n,m}$ is a Gaussian random vector with zero mean and covariance matrix  $\mathbf{Q}_{k,n,m} $, where
$\mathbf{Q}_{k,n,m} \in {\mathbb C}^{(r+1)\times(r+1)}$ is a matrix with entries $[\mathbf{Q}_{k,n,m}]_{\alpha\beta}={ E}_{  [ \mathbf{W}_k]_{n,m} }
\left[\left(v_{k,n,m}^{(\alpha)}\right)^H v_{k,n,m}^{(\beta)} \right]=g_{k,n,m} \left(\mathbf{x}_k^{(\alpha)}\right)^H \mathbf{u}_{{\rm T}_k,m}  \mathbf{u}_{{\rm
T}_k,m}^{H} \mathbf{x}_k^{(\beta)}$ {\bl for} $ \alpha =0,1,\cdots,r$, $\beta = 0,1,\cdots,r$. For ease of notation, we further define $\mathbf{T}_{k,m} =
\mathbf{u}_{{\rm T}_k,m} \mathbf{u}_{{\rm T}_k,m}^{H}$ and $\mathbf{R}_{k,n} = \mathbf{u}_{{\rm R}_k,n}\mathbf{u}_{{\rm R}_k,n}^{H}$. Therefore, we have
$[\mathbf{Q}_{k,n,m}]_{\alpha\beta} = g_{k,n,m} \left(\mathbf{x}_k^{(\alpha)}\right)^H \mathbf{T}_{k,m}\mathbf{x}_k^{(\beta)}$. Using (\ref{eq:HforV}) and letting
${\mathbb Q} = \{\mathbf{Q}_{k,n,m}\}_{\forall k,n,m}$, where ${\forall k,n,m}$ stands for $k = 1,2,\cdots,K$, $m = 1,2,\cdots,N_t$, and $n = 1,2,\cdots,N_r$, we
have
\begin{equation} \label{eq:ap_sf_E2-1}
   { E}_{\mathbf{H}}\left[\int D \mathbf{y} \prod_{\alpha=0}^re^{- \left\|{\bf y}-\sum_{k=1}^{K} \mathbf{H}_k \mathbf{x}_k^{(\alpha)}\right\|^2}\right]
    = e^{ \mathcal{S}({\mathbb Q})}
\end{equation}
where
\begin{multline}\label{eq:G1}
 \mathcal{S}({\mathbb Q})  =  \ln \int D \mathbf{y}  \\
 \times { E}_{\mathbf{V}}\left[\!\prod_{\alpha=0}^{r}e^{-\left\|{\bf y}- \sum_{k=1}^{K} \left(\sum_{n=1}^{N_r} \left(\sum_{m=1}^{N_t} v_{k,n,m}^{(\alpha)} \right) \mathbf{u}_{{\rm R}_k,n} \right) \right\|^2} \right].
 \end{multline}
Clearly, the interactions between $\mathbf{H}$ and $\mathbf{X}$ depend only on ${\mathbb Q}$. Therefore, it is useful to separate the expectation over $\mathbf{X}$
in (\ref{eq:ap_sf_E1}) into an integral over all possible $\mathbf{Q}_{k,n,m}$ and all possible $\mathbf{x}_k^{(\alpha)}$ configurations for a given
$\mathbf{Q}_{k,n,m}$ by introducing a $\delta$-function,
\begin{multline} \label{eq:ap_sf_E2-1}
 { E}_{{\bf y},\mathbf{H}}\left[\left(Z({\bf y},\mathbf{H})\right)^r \right]
    = \\ { E}_{\mathbf{X}}\left[ \int \prod_{k,n,m}\prod_{0\leq\alpha\leq\beta}^{r} d[\mathbf{Q}_{k,n,m}]_{\alpha\beta} \,
    e^{ \mathcal{S}({\mathbb Q})} \right. \\ \times
    \prod_{k,n,m}\prod_{0\leq\alpha\leq\beta}^{r}\delta\left(g_{k,n,m}  \left(\mathbf{x}_k^{(\alpha)}\right)^H \mathbf{T}_{k,m} \mathbf{x}_k^{(\beta)} \right. \\
\left. \left. - [\mathbf{Q}_{k,n,m}]_{\alpha\beta} \vphantom{g_{k,n,m}  \left(\mathbf{x}_k^{(\alpha)}\right)^H \mathbf{T}_{k,m} \mathbf{x}_k^{(\beta)}} \right)
  \vphantom{\int \prod_{k,n,m}}  \right].
\end{multline}
Let
\begin{multline}\label{eq:G2}
 \mu({\mathbb Q}) = { E}_{\mathbf{X}}\left[\prod_{k,n,m}\prod_{0\leq\alpha\leq\beta}^{r}\delta\left(g_{k,n,m}  \left(\mathbf{x}_k^{(\alpha)}\right)^H \mathbf{T}_{k,m} \mathbf{x}_k^{(\beta)} \right. \right. \\
\left. \left. - [\mathbf{Q}_{k,n,m}]_{\alpha\beta} \vphantom{g_{k,n,m}  \left(\mathbf{x}_k^{(\alpha)}\right)^H \mathbf{T}_{k,m} \mathbf{x}_k^{(\beta)}} \right)
\vphantom{\prod_{k,n,m}\prod_{0\leq\alpha\leq\beta}^{r}} \right].
 \end{multline}
Clearly, (\ref{eq:ap_sf_E2-1}) can be written as
\begin{equation} \label{eq:ap_sf_E2}
    { E}_{{\bf y},\mathbf{H}}\left[\left(Z({\bf y},\mathbf{H})\right)^r \right]
    = \int  e^{ \mathcal{S}({\mathbb Q})} d\mu({\mathbb Q}).
\end{equation}

Now, integrating the function in (\ref{eq:ap_sf_E2-1}) over ${\bf y}$, (\ref{eq:G1}) becomes
\begin{equation}\label{eq:G3}
    e^{\mathcal{S}({\mathbb Q})}={ E}_{\mathbf{V}}\left[\frac{1}{(r+1)^{N}}e^{- {\rm tr}\left( \left(\sum_{k}(\mathbf{U}_{{\rm R}_k}\mathbf{V}_k ) \right) {\bf \Sigma} \left(\sum_{k}(\mathbf{U}_{{\rm R}_k}\mathbf{V}_k ) \right)^{H} \right)}\right]
\end{equation}
where ${\bf \Sigma} = -\frac{1}{(r+1)}{\bf 1}{\bf 1}^H+\mathbf{I}_{r + 1}$. {\bl Recalling that $\mathbf{v}_{k,n,m}$ is a zero-mean Gaussian vector with covariance
matrix $\mathbf{Q}_{k,n,m} $, we obtain that $\sum_{k}(\mathbf{U}_{{\rm R}_k}\mathbf{V}_k)$ is a zero-mean Gaussian random vector with covariance $\mathbf{Q} \otimes
\mathbf{R} = \sum_{k,n}\left(\sum_{m} \mathbf{Q}_{k,n,m}\right) \otimes \mathbf{R}_{k,n} \in {\mathbb C}^{(r+1)N_r\times(r+1)N_r}$.} Thus, applying Lemma
\ref{Lemma_2}, we can eliminate $\mathbf{V}$ resulting in
\begin{equation}\label{eq:G4}
    e^{\mathcal{S}({\mathbb Q})}=\frac{1}{(r+1)^{N}\det\left(\mathbf{I}_{(r+1)N_r} + \mathbf{Q} {\boldsymbol \Sigma} \otimes \mathbf{R} \right)}.
\end{equation}
Inserting (\ref{eq:G4}) into (\ref{eq:ap_sf_E2}), we then deal with the expectation over $\mathbf{X}$ for a given ${\mathbb Q}$. Notice that only
 $\mu({\mathbb Q})$ is related to the components of $\mathbf{X}$. Using the inverse Laplace transform of the
$\delta$-function\footnote{The inverse Laplace transform of $\delta$-function is given by \cite{Nishimori2001} \begin{equation*} \delta(x)=\frac{1}{2\pi j}
\int_{-j\infty + t}^{j\infty + t} e^{\tilde{Q} x} d \tilde{Q},~\forall t \in {\mathbb R}.
\end{equation*}
}, $\mu({\mathbb Q})$ can be written as an exponential representation with integrals over auxiliary variables $\{\tilde{Q}_{k,n,m}^{(\alpha,\beta)}\}$. For ease of
notation, let $\tilde{\mathbf{Q}}_{k,n,m}\in {\mathbb C}^{(r+1)\times(r+1)}$ be a Hermitian matrix whose elements are the auxiliary variables
$\{\tilde{Q}_{k,n,m}^{(\alpha,\beta)}\}$. Similar to the definition of ${\mathbb Q}$, we further define the set $\tilde{{\mathbb Q}}
=\{\tilde{\mathbf{Q}}_{k,n,m}\}_{\forall k,n,m}$. In the large dimensional limit, the integrals over $\{\tilde{\mathbf{Q}}_{k,n,m}\}$ can be performed by
maximizing the exponent in $ \mu({\mathbb Q})$ with respect to $\{\tilde{\mathbf{Q}}_{k,n,m}\}$ (the saddle point method). Using the saddle point method and
following a similar approach as in \cite{wen2007asymptotic}, we can show that if $N_r$ is large, then $ \mu({\mathbb Q})$ is dominated by the exponent
\begin{multline}\label{eq:defJ}
    \mathcal{J}({\mathbb Q}) = \max_{\tilde{{\mathbb Q}}}  \left\{ \sum_{k,n,m} \tr \left(\tilde{\mathbf{Q}}_{k,n,m}\mathbf{Q}_{k,n,m}\right) \right. \\
  \left.   -\ln{ E}_{\mathbf{X}}\left[e^{\sum_{k,m}\tr \left( \sum_{n} g_{k,n,m} \tilde{\mathbf{Q}}_{k,n,m}\mathbf{X}_k^H \mathbf{T}_{k,m}\mathbf{X}_k\right)  }\right] \vphantom{\sum_{k,n,m} \tr \left(\tilde{\mathbf{Q}}_{k,n,m}\mathbf{Q}_{k,n,m}\right)} \right\}.
\end{multline}
{\bl
Similarly, by applying the saddle point method to (\ref{eq:ap_sf_E2}), we have \cite{Tanaka2002TIT,Guo2005TIT}
\begin{equation}\label{eq:calF1}
 - \ln { E}_{{\bf y},\mathbf{H}}\left[\left(Z ({\bf y},\mathbf{H})\right)^r\right]
 \simeq - \max_{{\mathbb Q}}\left\{{\cal S}({\mathbb Q})- {\cal J}({\mathbb Q}) \right\}
 = {\cal F} .
\end{equation}

The extremum over $\tilde{{\mathbb Q}}$ and ${\mathbb Q}$ in (\ref{eq:defJ}) and (\ref{eq:calF1}) can be obtained via {seeking the point of zero gradient with
respect to $\tilde{{\mathbb Q}}$ and ${\mathbb Q}$, respectively,} yielding a set of self-consistent equations. To avoid searching for the {extremum} over all
possible ${\mathbb Q}$ and $\tilde{{\mathbb Q}}$, we make the following {\it replica symmetry} (RS) assumption for the saddle point:
\begin{align}
 \mathbf{Q}_{k,n,m} &=q_{k,n,m}{\bf 11}^H +(c_{k,n,m}-q_{k,n,m})\mathbf{I}_{r+1} \\
\tilde{\mathbf{Q}}_{k,n,m} &=\tilde{q}_{k,n,m}{\bf 11}^H+(\tilde{c}_{k,n,m}-\tilde{q}_{k,n,m})\mathbf{I}_{r+1}.
\end{align}
 {\bl With this RS assumption, the problem of seeking the extremum in (\ref{eq:calF1}) with respect to $(\mathbf{Q}_{k,n,m},\tilde{\mathbf{Q}}_{k,n,m})$ is reduced
to seeking the extremum over the four parameters $(q_{k,n,m},c_{k,n,m},\tilde{q}_{k,n,m},\tilde{c}_{k,n,m})$. Although the RS assumption is heuristic, and cases of RS breaking appear in literature \cite{Nishimori2001,Zaidel2012TIT}, it is widely used in physics \cite{Nishimori2001} and information theory
\cite{wen2007asymptotic,Tanaka2002TIT,RMuller2008JSAC,Guo2005TIT,wen2011sum,Mou-03,Zaidel2012TIT}.}  {\bfbl Also, some results obtained based on the RS assumption
have been shown to become exact in the large system limit \cite{Bayati2011TIT}. }

Based on the RS assumption, $\mathcal{S}({\mathbb Q})$ becomes
\begin{multline}\label{eq:RS_G}
\mathcal{S}({\mathbb Q})
 = - N_r \ln(r+1)  \\ - r \ln \det\left(\mathbf{I}_{N_r} +\sum_{k,n} \left(\sum_{m}c_{k,n,m}-q_{k,n,m}\right) \mathbf{R}_{k,n} \right) .
\end{multline}
Applying Lemma \ref{Lemma_3}, one can easily show that
\begin{multline} \label{eq:RSQQ}
 \sum_{k,n,m}{\rm tr} \left(\tilde{\mathbf{Q}}_{k,n,m}\mathbf{Q}_{k,n,m}\right)
 =  \\ \sum_{k,n,m} (\tilde{c}_{k,n,m} +r\tilde{q}_{k,n,m})(c_{k,n,m} +rq_{k,n,m}) \\ + r (\tilde{c}_{k,n,m} - \tilde{q}_{k,n,m})(c_{k,n,m} - q_{k,n,m}).
\end{multline}
Therefore, the last term of (\ref{eq:defJ}) can be written as
\begin{multline} \label{eq:LT_Rate_Fun1}
 \ln{ E}_{{\bf X}}\left[e^{ \sum_{k,m}{ \tr}\left( \sum_{n} g_{k,n,m} \tilde{\mathbf{Q}}_{k,n,m}\mathbf{X}_k^H \mathbf{T}_{k,m}\mathbf{X}_k\right)}  \right] \\
 =  \ln { E}_{{\bf X}}\left[e^{ {\rm vec}(\mathbf{X})^H \tilde{\mathbf{T}} {\rm vec}(\mathbf{X})}\right]
\end{multline}
where we have used $\tilde{\mathbf{T}} = {\rm diag}\left(\tilde{\mathbf{T}}_1,\tilde{\mathbf{T}}_2, \dots,\tilde{\mathbf{T}}_K \right)$ and $\tilde{\mathbf{T}}_k =
\sum_{m} \Big( \left(\sum_{n} g_{k,n,m} \tilde{\mathbf{Q}}_{k,n,m}\right) \otimes \mathbf{T}_{k,m} \Big)$.  For ease of notation, we define $\boldsymbol{\Xi}'  =
\mathbf{T}'(0)$ and $ \boldsymbol{\Xi}  = \mathbf{T}'(-1)$, where $\mathbf{T}'(\tau) = {\rm{blockdiag}} \left(\mathbf{T}'_1(\tau), \mathbf{T}'_2(\tau),
\dots,\mathbf{T}'_K(\tau)\right)$ and $\mathbf{T}'_k(\tau) = \sum_{k,m} \left(\sum_{n} g_{k,n,m}(\tau\tilde{c}_{k,n,m} + \tilde{q}_{k,n,m}) \right)
\mathbf{T}_{k,m}$. Using the above definitions in (\ref{eq:LT_Rate_Fun1}) yields
\begin{equation} \label{eq:LT_Rate_Fun2}
\begin{array}{l}
\ln { E}_{{\bf X}}\left[e^{ {\rm vec}(\mathbf{X})^H \tilde{\mathbf{T}} {\rm vec}(\mathbf{X})}\right] = \\
\hspace{-0.2cm} \ln { E}_{{\bf X}}\left[e^{ \left(\sum_{\alpha=0}^{r} \sqrt{\boldsymbol{\Xi}'} \mathbf{x}^{(\alpha)} \right)^{H}\left(\sum_{\alpha=0}^{r} \sqrt{\boldsymbol{\Xi}'} \mathbf{x}^{(\alpha)} \right)
 - \sum_{\alpha=0}^{r} (\mathbf{x})^{(\alpha)^H} \boldsymbol {\Xi} \mathbf{x}^{(\alpha)} }\right].
 \end{array}
\end{equation}
Now, we decouple the first quadratic term in the exponent of (\ref{eq:LT_Rate_Fun2}) by using the Hubbard-Stratonovich transformation
in Lemma \ref{Lemma_4} and introducing the auxiliary vector $\mathbf{z}$. As a result,
(\ref{eq:LT_Rate_Fun2}) becomes
\begin{align} \label{eq:LT_Rate_Fun3}
 &\ln  \int D\mathbf{z} ~{E}_{{\bf X}}\left[e^{ - g(\mathbf{z}) }\right]
\end{align}
where
\begin{multline} \label{g(z)}
g(\mathbf{z}) = \mathbf{z}^{H}\mathbf{z} + \left(\sum_{\alpha=0}^{r} \sqrt{\boldsymbol{\Xi}'} \mathbf{x}^{(\alpha)} \right)^{H}\mathbf{z} \\
 + \mathbf{z}^{H}\left(\sum_{\alpha=0}^{r} \sqrt{\boldsymbol{\Xi}'} \mathbf{x}^{(\alpha)} \right)
- \sum_{\alpha=0}^{r} (\mathbf{x}^{(\alpha)})^H \boldsymbol{\Xi} \mathbf{x}^{(\alpha)}.
\end{multline}
Inserting (\ref{eq:RS_G}), (\ref{eq:RSQQ}), and (\ref{eq:LT_Rate_Fun3}) into (\ref{eq:calF1}), we obtain ${\cal F}$ under the RS assumption as
\begin{equation} \label{eq:midFree}
    {\cal F} = - \max_{\mathbf{c},\mathbf{q}}\min_{\tilde{\mathbf{c}},\tilde{\mathbf{q}}}{\cal T}^{(r)}(\mathbf{c},\mathbf{q},\tilde{\mathbf{c}},\tilde{\mathbf{q}})
\end{equation}
where
\begin{multline} \label{eq:TreplicaSym}
    -{\cal T}^{(r)}(\mathbf{c},\mathbf{q},\tilde{\mathbf{c}},\tilde{\mathbf{q}})
    =~ \int D\mathbf{z} {E}_{\mathbf{X}}\left[e^{ -\left\| \mathbf{z} - \sqrt{\boldsymbol{\Xi}'}\mathbf{x} \right\|^2 + \mathbf{x}^{H}(\boldsymbol{\Xi}'-\boldsymbol{\Xi})\mathbf{x} }\right] \\
   \times \left( {E}_{\mathbf{X}}\left[e^{ \left( \sqrt{\boldsymbol{\Xi}'}\mathbf{x} \right)^{H}\mathbf{z}
    + \mathbf{z}^{H}\left( \sqrt{\boldsymbol{\Xi}'}\mathbf{x} \right) - \mathbf{x}^{H} \boldsymbol{\Xi} \mathbf{x} } \right] \right)^r  + N_r\ln(r+1) \\
     +r\ln\det\left(\mathbf{I}_{N_r}+\sum_{k,n} \left(\sum_{m}c_{k,n,m}-q_{k,n,m}\right) \mathbf{R}_{k,n} \right)   \\
 +\sum_{k,n,m} (\tilde{c}_{k,n,m} +r\tilde{q}_{k,n,m})(c_{k,n,m} +rq_{k,n,m}) \\
 + r (\tilde{c}_{k,n,m} - \tilde{q}_{k,n,m})(c_{k,n,m} -q_{k,n,m}).
\end{multline}
The parameters $\{c_{k,n,m},q_{k,n,m},\tilde{c}_{k,n,m},\tilde{q}_{k,n,m}\}$ are determined by seeking the point of zero gradient with respect to
$\{c_{k,n,m},q_{k,n,m},\tilde{c}_{k,n,m},\tilde{q}_{k,n,m}\}$. It is easy to check that $\tilde{c}_{k,n,m} = 0, ~\forall k, n,m$ and $c_{k,n,m} = \tr
(\mathbf{T}_{k,m}), ~\forall k,n,m$.

Motivated by the exponent of the first term on the right hand side of (\ref{eq:TreplicaSym}), we define a virtual MIMO channel as in (\ref{eq:EqScalGAUEach}),
where $\mathbf{z}_{\cal_{A}} := \mathbf{z}$, $\mathbf{T}_{\cal A} := \boldsymbol{\Xi}'$, and $\mathbf{B}_{\cal A} \mathbf{d}_{\cal A} := \mathbf{x}$. This virtual
MIMO channel does not relate to any physical channel model and is introduced only for clarity of notation. In particular, we will show that the first term on the
right hand side of (\ref{eq:TreplicaSym}) can be written as the mutual information of the virtual MIMO channel when taking the derivative of ${\cal T}^{(r)}$ with
respect to $r$ at $r=0$. Recall from {\bl (\ref{eq:ap_sf_F}) and (\ref{eq:calF1}) that} we are only interested in the derivative of ${\cal F}$ at $r=0$. Let
$\gamma_{k,n,m} = \tilde{q}_{k,n,m}$ and $\psi_{k,n,m} = c_{k,n,m}-q_{k,n,m}$. Hence, from (\ref{eq:midFree}), the final result can be expressed as
\begin{align}
 F &\simeq \! -\log_2 e \lim_{r\rightarrow 0}\frac{\partial}{\partial r} {\cal F} \nonumber \\
   & = \ I\left( \mathbf{x};\mathbf{z}
 \big| \sqrt{\boldsymbol \Xi} \right) \! + \! \log_2\det\left(\mathbf{I}_{N_r}+\mathbf{R} \right) \nonumber \\
 & \qquad\qquad -{\log_2 e} \sum_{k,n,m} \gamma_{k,n,m} \psi_{k,n,m} + {\log_2 e} N_r  \label{eq:ap_GenFree}
\end{align}
where $\boldsymbol{\Xi} = \mathbf{T}$, $\mathbf{T} =  {\rm{blockdiag}} \left(\mathbf{T}_1,\mathbf{T}_2, \dots,\mathbf{T}_K\right)$, $\mathbf{T}_k = \sum_{m}
\left(\sum_{n} g_{k,n,m} \gamma_{k,n,m} \right)\mathbf{T}_{k,m}$ and $\mathbf{R} = \sum_{k,n} \left(\sum_{m} \psi_{k,n,m}\right)$ $\mathbf{R}_{k,n}$. The
parameters $\gamma_{k,n,m}$ and $\psi_{k,n,m}$ are determined by seeking the point of zero gradient of $F$ with respect to $\psi_{k,n,m}$ and $\gamma_{k,n,m}$,
respectively. Hence, we have
\begin{align} \label{eq:ap_gamma}
\gamma_{k,n,m} &= \tr{\Big( \left(\mathbf{I}_{N_r}+\mathbf{R}\right)^{-1} \mathbf{R}_{k,n} \Big)}
\end{align}
and
\begin{align}\label{eq:ap_fai}
\psi_{k,n,m} &= \ln 2 \ \frac{\partial}{\partial \gamma_{k,n,m}}
 I\left(\mathbf{x};\mathbf{z} \left|\sqrt{\boldsymbol \Xi}\right.\right)
 = g_{k,n,m} \tr \left( \boldsymbol{\Omega}_k \mathbf{T}_{k,m} \right)
\end{align}
where the derivative of the mutual information follows from the relationship between the mutual information and the MMSE revealed in \cite{Palomar2006TIT}. Let $\gamma_{k,n} = \gamma_{k,n,m}$ and $\psi_{k,m} = \tr \left( \boldsymbol{\Omega}_k \mathbf{T}_{k,m} \right)$, for $m=1,2,\dots,M$.
Using (\ref{eq:ap_GenFree}) and substituting the
definitions of $\gamma_{k,n}$, $\psi_{k,m}$, $\mathbf{T}_{\cal A}$, $\mathbf{B}_{\cal A}$, and model
(\ref{eq:EqScalGAUEach}), we  obtain (\ref{eq:GAUMutuall_2}) for the case $K_1 = K$.
The case for arbitrary values of $K_1$ can be proved following a similar approach as above.

{\bl
\section{Proof of Theorem \ref{theo:opt_structure}}\label{sec:proof_opt_structure}
Using the notations introduced in Section III-B and according to Proposition \ref{prop:ach_rate_mac}, we obtain an asymptotic expression for
$I(\mathbf{d}_1,\cdots, \mathbf{d}_k ;  \mathbf{y}| \mathbf{d}_{k+1},\cdots,\mathbf{d}_{K})$ as
\begin{multline}\label{eq:GAUMutuall_2_nec}
I\left( {{\bf{d}}_{{\cal A}_k} ;{\bf{y}}\left| {{\bf{d}}_{{{\cal A}_k^c} } } \right.} \right) \simeq
\sum\limits_{t = 1}^{k} {I\left( {{\bf{d}}_{t} ;{\bf{z}}^{(k)}_{t} \left| {\sqrt {{\bf{T}}^{(k)}_{t} } {\bf{B}}_{t} } \right.} \right)} \\
 + \log_2 \det\left(\mathbf{I}_{N_r} +\mathbf{R}_{{\cal A}_k}\right) - \log_2 e \sum_{t=1}^{k} \left(\boldsymbol{\gamma}^{(k)}_{t}\right)^T \mathbf{G}_{t} \boldsymbol{\psi}^{(k)}_{t}.
\end{multline}

To investigate the optimal precoders which maximize $R_{\rm sum,asy}^{\bl {\rm w}} \left({\mathbf{B}}_1 ,{\mathbf{B}}_2 , \cdots, {\mathbf{B}}_K \right)$, we
consider the gradient of $R_{\rm sum,asy}^{\bl {\rm w}} \left({\mathbf{B}}_1 ,{\mathbf{B}}_2 , \cdots, {\mathbf{B}}_K \right)$ with respect to $\mathbf{B}_l$. It is noted from (\ref{eq:eqChMatrixTR})--(\ref{eq:obj_wsr_2}) that the parameters $I\left( {{\bf{d}}_{t}
;{\bf{z}}^{(k)}_{t} \left| {\sqrt {{\bf{T}}^{(k)}_{t} } {\bf{B}}_{t} } \right.} \right)$, ${{\gamma}}^{(k)}_{t,n}$, ${{\psi}}^{(k)}_{t,m}$ depend on
$\mathbf{B}_l$. Therefore, the gradient of $R_{\rm sum,asy}^{\bl {\rm w}} \left({\mathbf{B}}_1 ,{\mathbf{B}}_2 , \cdots, {\mathbf{B}}_K \right)$ with respect to
$\mathbf{B}_l$ is given by
\begin{multline}\label{eq:GAUMutuall_2_grad}
\begin{array}{l}
\sum\limits_{k = 1}^K {{\Delta }}_k  \sum\limits_{t = 1}^{k} \nabla_{\mathbf{B}_l} {I\left( {{\bf{d}}_{t} ;{\bf{z}}^{(k)}_{t} \left| {\sqrt {  {\bf{T}}^{(k)}_{t} } {\bf{B}}_{t} } \right.} \right)} \\
\hspace{-0.5cm}  +  \!  \log_2 e \! \!  \sum\limits_{k = 1}^K {{\Delta }}_k \! \sum\limits_{t=1}^{k} \! \sum\limits_{n=1}^{N_r} \! \frac{{\partial R_{{\rm{sum,asy}}}^{\rm{w}}\left( {{{\bf{B}}_1},{{\bf{B}}_2}, \cdots ,{{\bf{B}}_K}} \right)}}{{\partial {\gamma}_{t,n}^{\left(k\right)}}} \!\nabla_{\mathbf{B}_l} {\gamma}^{(k)}_{t,n}(\mathbf{B}_l)  \\
 \hspace{-0.5cm} + \!   \log_2 e \! \!  \sum\limits_{k = 1}^K {{\Delta }}_k \! \sum\limits_{t=1}^{k} \!\sum\limits_{m=1}^{N_t} \!\frac{{\partial R_{{\rm{sum,asy}}}^{\rm{w}}\left( {{{\bf{B}}_1},{{\bf{B}}_2}, \cdots ,{{\bf{B}}_K}} \right)}}{{\partial {\psi}^{(k)}_{t,m} }} \! \nabla_{\mathbf{B}_l} {\psi}^{(k)}_{t,m}(\mathbf{B}_l) \\
 \hspace{2cm}  \ \ l = 1,2,\cdots, K.
\end{array}
\end{multline}

Based on the definition of ${{\gamma}}^{(k)}_{t,n}$ and
${{\psi}}^{(k)}_{t,m}$  in Appendix \ref{sec:proof_ach_rate_mac},
we know that ${{\gamma}}^{(k)}_{t,n}$ and ${{\psi}}^{(k)}_{t,m}$ are obtained by setting the partial
derivatives of the asymptotic mutual information expression in Proposition \ref{prop:ach_rate_mac} with
respective to ${{\gamma}}^{(k)}_{t,n}$ and ${{\psi}}^{(k)}_{t,m}$ to zero. Then, according to
the definition of $R_{{\rm{sum,asy}}}^{\rm{w}}$ in (\ref{eq:obj_wsr_3}), we obtain
\begin{align} \label{eq:der_gamma_psi}
\frac{{\partial R_{{\rm{sum,asy}}}^{\rm{w}}\left( {{{\bf{B}}_1},{{\bf{B}}_2}, \cdots ,{{\bf{B}}_K}} \right)}}{{\partial {\gamma}_{t,n}^{\left(k\right)}}} = 0 \\
 \frac{{\partial R_{{\rm{sum,asy}}}^{\rm{w}}\left( {{{\bf{B}}_1},{{\bf{B}}_2}, \cdots ,{{\bf{B}}_K}} \right)}}{{\partial {\psi}^{(k)}_{t,m} }} =0 .
\end{align}
As a result, we have
\begin{multline} \label{eq:grad_B_l}
{\nabla _{{{\bf{B}}_l}}}R_{{\rm{sum,asy}}}^{\rm{w}}\left( {{{\bf{B}}_1},{{\bf{B}}_2}, \cdots ,{{\bf{B}}_K}} \right) \\
 = \sum\limits_{k = l}^K {{\Delta _k}{\nabla _{{{\bf{B}}_l}}}I\left( {{{\bf{d}}_l};{\bf{z}}_l^{\left( k \right)}\left| {\sqrt {{\bf{T}}_l^{\left( k \right)}} {{\bf{B}}_l}} \right.} \right)}.
\end{multline}
% Thus, for the asymptotic WSR maximization, we may consider ${{\gamma}}^{(k)}_{t,n}$  and ${{\psi}}^{(k)}_{t,m}$ as being independent of $\mathbf{B}_l$, $l = 1,2,\cdots,K$.
From (\ref{eq:grad_B_l}), we know that the asymptotic WSR maximization problem is equivalent to the following $K$ subproblems:
\begin{equation} \label{eq:subproblem}
\begin{aligned}
 \max_{ \mathbf{B}_{l} } & ~ \sum\limits_{k = l}^K {{\Delta _k}I\left( {{{\bf{d}}_l};{\bf{z}}_l^{\left( k \right)}\left| {\sqrt {{\bf{T}}_l^{\left( k \right)}} {{\bf{B}}_l}} \right.} \right)},\\
 \text{s.t.} & ~\text{tr} \left(\mathbf{B}_{l}\mathbf{B}^H_{l}\right) \leq P_{l} ,
\end{aligned}
\end{equation}
%\begin{equation}\label{eq:subproblem_constraint}
%       \text{tr} \left(\mathbf{B}_{l}\mathbf{B}^H_{l}\right) \leq P_{l}
%\end{equation}
where $l = 1,2,\cdots,K$.

From (\ref{eq:mutual_info}), we have
\begin{align}
&I\left( {{\bf{d}}_{l} ;{\bf{z}}_{l}^{(k)} \left| {\sqrt {{\bf{T}}_{l}^{(k)} } {\bf{B}}_{l} } \right.} \right) = \log _2 M_{l}  - \frac{1}{{M_{l} }} \nonumber \\
& \times \sum\limits_{m = 1}^{M_{l } } {E_{{\bf{v}} } \! \left\{ \!  {\log _2 \sum\limits_{p = 1}^{M_{l } } {e^{ - \left( {\left\| {\sqrt {{\bf{T}}_{l}^{(k)} } {\bf{B}}_{l} \left( {{\bf{a}}_{l,p} \!  - \! {\bf{a}}_{l,m} } \right) + {\bf{v}} } \right\|^2  - \left\| {{\bf{v}} } \right\|^2 } \right)} } } \! \right\}} \nonumber \\ % \label{eq:mutual_info_sub} \\
& = {\log _2}{M_l} - \frac{1}{{{M_l}}}\sum\limits_{m = 1}^{{M_l}} {{E_{\bf{v}}}} \left\{ {{{\log }_2}\sum\limits_{p = 1}^{{M_l}} {{e^{ - {l_{pm,lk}}\left( {\bf{v}} \right)}}} } \right\}\label{eq:mutual_info_sub_2}
\end{align}
where
\begin{multline}\label{eq:l_pmlk}
{l_{pm,lk}}\left( {\bf{v}} \right) = \tr\left( {\left( {{{\bf{a}}_{l,p}} - {{\bf{a}}_{l,m}}} \right){{\left( {{{\bf{a}}_{l,p}} - {{\bf{a}}_{l,m}}} \right)}^H}{\bf{B}}_l^H{\bf{T}}_l^{\left( k \right)}{{\bf{B}}_l}} \right)  \\
+ 2{\mathop{\rm Re}\nolimits} \left( {{{\bf{v}}^H}\sqrt {{\bf{T}}_l^{\left( k \right)}} {{\bf{B}}_l}\left( {{{\bf{a}}_{l,p}} - {{\bf{a}}_{l,m}}} \right)} \right).
\end{multline}

Based on (\ref{eq:mutual_info_sub_2}), the value of $I\left( {{\bf{d}}_{l} ;{\bf{z}}_{l}^{(k)} \left| {\sqrt {{\bf{T}}_{l}^{(k)} } {\bf{B}}_{l} } \right.} \right)$
is determined by  matrix $\mathbf{Q}_l = {\bf{B}}_l^H{\bf{T}}_l^{\left( k \right)}{{\bf{B}}_l}$.  Define the eigenvalue decomposition of $\mathbf{Q}_l$ as
${{\bf{Q}}_l} = {{\bf{U}}_{q,l}}{\boldsymbol{\Gamma}_{q,l}}{\bf{U}}_{q,l}^H$. Then, we have
\begin{align}\label{eq:dec_sub}
{\left( {{\bf{T}}_l^{\left( k \right)}} \right)^{1/2}}{{\bf{B}}_l} = {{\bf{V}}_{q,l}}{ {{\boldsymbol{\Gamma} _{q,l}^{1/2}}}}{\bf{U}}_{q,l}^H
\end{align}
where ${{\bf{V}}_{q,l}} \in \mathbb{C}^{N_t \times N_t}$ is an arbitrary unitary matrix. According to (\ref{eq:eqChMatrixTR}) and (\ref{eq:dec_sub}), ${{\bf{B}}_l}$ can be
expressed as
\begin{align}\label{eq:B_sub}
{{\bf{B}}_l} & = {\left( {{\bf{T}}_l^{\left( k \right)}} \right)^{ - 1/2}}{{\bf{V}}_{q,l}}{ {{\boldsymbol{\Gamma} _{q,l}^{1/2}}}}{\bf{U}}_{q,l}^H \\
& = {{\bf{U}}_{{{\rm{T}}_l}}}{\rm diag}{\left( {{\bf{G}}_l^T{\bl \boldsymbol{\gamma}}_l^{\left( k \right)}} \right)^{ - 1/2}}{\bf{U}}_{{{\rm{T}}_l}}^H{{\bf{V}}_{q,l}}{{{\boldsymbol{\Gamma} _{q,l}^{1/2}}} }{\bf{U}}_{q,l}^H . \label{eq:B_sub_2}
\end{align}

Any ${{\bf{B}}_l}$ which conforms to the expression in (\ref{eq:B_sub_2}) satisfies ${\bf{B}}_l^H{\bf{T}}_l^{\left( k \right)}{{\bf{B}}_l} = \mathbf{Q}_l$.
Furthermore, the transmit power of ${{\bf{B}}_l}$ is given by
\begin{align}\label{eq:power_sub}
\tr\left( {{{\bf{B}}_l}{\bf{B}}_l^H} \right) & = \tr\left( { {\rm diag}{\left( {{\bf{G}}_l^T{\bl \boldsymbol{\gamma}}_l^{\left( k \right)}} \right)^{ - 1}
{\bf{U}}_{{{\rm{T}}_l}}^H{{\bf{V}}_{q,l}}{ \boldsymbol{\Gamma}_{q,l}}{\bf{V}}_{q,l}^H{{\bf{U}}_{{{\rm{T}}_l}}}} }\right) \\
& \mathop  \ge \limits^{\left( a \right)} \tr\left( {  {\rm diag}{\left( {{\bf{G}}_l^T{\bl \boldsymbol{\gamma}}_l^{\left( k \right)}} \right)^{ - 1} {\boldsymbol{\Gamma} _{q,l}}}} \right) \label{eq:power_sub_2}
\end{align}
where (a) is obtained based on \cite[Eq. (3.158)]{Palomar2006}. The equality in (\ref{eq:power_sub_2}) holds when
${\bf{V}}_{q,l}^H {{\bf{U}}_{{{\rm{T}}_l}} = \mathbf{I}_{N_t}}$. By substituting  this condition into (\ref{eq:B_sub_2}), we obtain
\begin{align}\label{eq:B_sub_3}
{{\bf{B}}_l} = {{\bf{U}}_{{{\rm{T}}_l}}}{\rm diag}{\left( {{\bf{G}}_l^T{\bl \boldsymbol{\gamma}}_l^{\left( k \right)}} \right)^{ - 1/2}}{{{\boldsymbol{\Gamma} _{q,l}^{1/2}}} }{\bf{U}}_{q,l}^H .
\end{align}
Eq. (\ref{eq:B_sub_3}) indicates that for $I\left( {{\bf{d}}_{l} ;{\bf{z}}_{l}^{(k)} \left| {\sqrt {{\bf{T}}_{l}^{(k)} } {\bf{B}}_{l} } \right.} \right)$,
setting the left singular matrix to be equal to ${{\bf{U}}_{{{\rm{T}}_l}}}$  minimizes the transmit power $\tr(\mathbf{B}_l \mathbf{B}_l^H)$.
This conclusion holds for all $k = l, l+ 1,\cdots,L$ in (\ref{eq:subproblem}).

We assume that the optimal precoder is $\mathbf{B}_l^{*}$, $l = 1,2,\cdots,K$.
 If the left singular matrix of $\mathbf{B}_l^{*}$ is not equal to ${{\bf{U}}_{{{\rm{T}}_l}}}$, we can always find another precoder with the left singular matrix equal to ${{\bf{U}}_{{{\rm{T}}_l}}}$, which achieves the same value of $\mathbf{Q}_l$ as $\mathbf{B}_l^{*}$. Therefore, this precoder can achieve the same mutual information  in (\ref{eq:subproblem}) as
 $\mathbf{B}_l^{*}$, but with a smaller transmit power according to (\ref{eq:power_sub_2}).
Then, by increasing the transmit power of this precoder until the power constraint is met with equality,  a larger mutual information value in (\ref{eq:subproblem}) is achieved. This contradicts the assumption that $\mathbf{B}_l^{*}$ is the optimal precoder.
As a result, the left singular  matrix of $\mathbf{B}_l^{*}$ must be equal to ${{\bf{U}}_{{{\rm{T}}_l}}}$. {\bfbl Substituting the optimal $\mathbf{B}_l$ in (\ref{eq:B_sub_3}) into (\ref{eq:obj_wsr_2}), we obtain (\ref{eq:subproblem_sim}). }
This completes the proof.

}


%\begin{figure}[!ht]
%\centering
%\includegraphics[width = 0.8\textwidth]{figure/bc_rx}
%\end{figure}
\end{document}